\documentclass[superscriptaddress,nofootinbib,amsmath,amssymb,showkeys]{revtex4-1}
\usepackage[utf8]{inputenc}
\pdfoutput=1
\usepackage{xcolor}
\usepackage{mathtools, amsmath,amsfonts,amssymb,amsthm, tabu, amsbsy}
\usepackage{xspace}
\usepackage{hyperref}
\usepackage{dsfont}
\usepackage{tikz}
\usetikzlibrary{arrows, decorations.markings, fit, shapes, patterns, hobby, calc,decorations.pathreplacing,positioning}
\usepackage{pgfplots}
\pgfplotsset{compat=1.9}
\usepackage{mathrsfs}
\usepackage{braket}
\usepackage{enumitem}
\usepackage[title]{appendix}
\usepackage{ifthen}
\usepackage{diagbox}
\usepackage{cases}
\usepackage{makecell}

\usepackage{interval}

\newcommand{\avgoe}[1]{\ensuremath{\left\langle #1\right\rangle}\xspace}
\newcommand{\Tr}[1]{\ensuremath{\textrm{Tr}\left(#1\right)}}
\newcommand{\tr}[1]{\ensuremath{\tau\left(#1\right)}}

\newcommand{\E}[1]{\ensuremath{\mathbb{E}\,\left[#1\right]}}
\newcommand{\X}{\ensuremath{\mathbf{X}}}
\renewcommand{\P}{\ensuremath{\mathbf{P}}}

\renewcommand{\i}{\ensuremath{\mathbf{i}}}
\renewcommand{\k}{\ensuremath{\mathbf{k}}}
\newcommand{\x}{\ensuremath{\mathbf{x}}}
\newcommand{\W}{\ensuremath{\mathbf{W}}}
\newcommand{\Xt}{\ensuremath{\mathbf{X}^\dagger}}
\newcommand{\eps}{\ensuremath{\epsilon}}
\newcommand{\goe}{{$\textsc{goe}$}}

\newcommand{\gee}{{$\textsc{gee}$}}
\newcommand{\iid}{{$\textsc{iid}$}}
\newcommand{\n}{\ensuremath{\boldsymbol{n}}}
\newcommand{\m}{\ensuremath{\boldsymbol{m}}}
\newcommand{\sd}[1]{\ensuremath{y^\star_{#1}}}
\newcommand{\mm}[2]{\ensuremath{\mathcal{O}_{#1}^{#2}}}
\newcommand{\rmm}[2]{\ensuremath{\widehat{\mathcal{O}}_{#1}^{#2}}}
\newcommand{\card}[1]{\ensuremath{\left|#1\right|}}
\DeclareMathOperator{\nc}{\text{NC}}

\newcommand{\secref}[1]{Section~\ref{#1}}
\newcommand{\figref}[1]{Fig.~\ref{#1}}
\newcommand{\tabref}[1]{Table~\ref{#1}}

\def \A{\mathbf{A}}
\def \B{\mathbf{B}}

\def \d{\mathrm{d}}

\def \N{\mathbb{N}}
\def \O{\mathbb{O}}

\def \R{\mathbb{R}}

\def\Ac{{\mathcal{A}}}

\def\Fc{{\mathcal{F}}}

\def\Oc{{\mathcal{O}}}

\def\Zc{{\mathcal{Z}}}

\newcommand{\Matching}[4]{%
    \foreach \x/\y/\z/\w in {#3} {
    {\ifthenelse{\z = \w}
        {\ifthenelse{\z = 0}
        {\draw(-\y+1,#4) to[bend left=60] (-\x+1,#4) ;}
        {\draw(\x,#4) to[bend left=60] (\y,#4);}}
        {\ifthenelse{\z = 0}
        {\draw(-\x+1,#4) to[bend left=60] (\y,#4) ;}
        {}}
    }
    }
    \foreach \x in {1,...,#1}{
       \draw[circle, opacity=1, fill=white] (-\x+1,#4)circle[radius=0.75mm]node[below]{${\x}$};
    }
    \foreach \x in {1,...,#2}{
       \draw[circle,fill] (\x,#4)circle[radius=0.75mm]node[below]{${\x}$};
    }
}
\colorlet{w}{white}
\colorlet{b}{black}

\tikzset{
  pics/carc/.style args={#1:#2:#3}{
    code={
      \draw[pic actions] (#1:#3) arc(#1:#2:#3);
    }
  }
}
\tikzset{
  ptm/.style n args={1}{circle,draw=black, fill={#1}, inner sep=0pt, outer sep=0pt, minimum size=1.5mm, scale=1},
  pt/.style n args={1}{insert path={node[circle,draw=black, fill=#1, inner sep=0pt, outer sep=0pt, minimum size=1.5mm] {}}},
  dnup/.style n args={3}{insert path={ [pt={#2}] .. controls +(0,1) and +(0,-1) .. +(#1,2)  [pt={#3}]}},
  dndn/.style n args={3}{insert path={ [pt={#2}] .. controls +(0,1) and +(0,1) .. +(#1,0) [pt={#3}]}},
  upup/.style n args={3}{insert path={ [pt={#2}] .. controls +(0,-1) and +(0,-1) .. +(#1,0) [pt={#3}]}},
  dnUp/.style n args={3}{insert path={ [pt={#2}] .. controls +(0,1) and +(0,-1) .. +(#1,4) [pt={#3}]}},
}

\newcommand{\CircleMatching}[4]{
\draw (#4) circle[radius=#3];
    \foreach \x/\y/\z/\w in {#2} {
        \node [ptm={\z}] (\x) at ($(360*\x/#1:#3)+(#4)$) {};
        \node [ptm={\w}] (\y) at ($(360*\y/#1:#3)+(#4)$) {};
        \draw (\x) to[bend left=100*(#1/2-(\y-\x))/#1]  (\y);
    }
}

\newcommand{\CircleMatchingNoOuter}[4]{
    \foreach \x/\y/\z/\w in {#2} {
        \node [ptm={\z}] (\x) at ($(360*\x/#1:#3)+(#4)$) {};
        \node [ptm={\w}] (\y) at ($(360*\y/#1:#3)+(#4)$) {};
        \draw (\x) to[bend left=100*(#1/2-(\y-\x))/#1]  (\y);
    }
}
\newcommand*{\myfont}{\fontfamily{lmss}\selectfont}
\DeclareTextFontCommand{\textlmss}{\myfont}
\newcommand{\Att}{\textlmss{A}}

\begin{document}

\title{Some Mixed-Moments of Gaussian Elliptic Matrices and Ginibre Matrices}
\author{Th\'{e}o Dessertaine}
\email{dessertainetheo@gmail.com}
\affiliation{LadHyX UMR CNRS 7646, Ecole polytechnique, 91128 Palaiseau Cedex, France}
\affiliation{Chair of Econophysics \& Complex Systems, Ecole polytechnique, 91128 Palaiseau Cedex, France}

\date{\today}
\begin{abstract}
We consider the mixed-moments $\varphi(\X^{\eps_1},\ldots,\X^{\eps_k})=\lim_{N\to\infty}N^{-1}\E{\Tr{\X^\eps_1\cdots\X^{\eps_k}}}$ of complex Gaussian Elliptic Matrices $\X$ (with correlation parameter $\rho$ between elements $\X_{ij}$ and $\X_{ji}^*$), where symbolically $\eps_i\in\{1,\dagger\}$, and where the expectation $\E{\cdot}$ is taken over all matrices $\X$. We start by finding an explicit formula for $\varphi(\X^n,(\Xt)^m)$, $n,m\in\N$, by using a mapping between non-crossing pairings on $\ell=n+m$ elements and Temperley-Lieb diagrams between two strands of $n$ and $m$ elements. This formula allows for a numerically efficient way to compute $\varphi(\X^n,(\Xt)^m)$ by reducing the exponential complexity of a naive enumeration of non-crossing pairings to polynomial complexity. We also provide the asymptotic behavior of these mixed-moments as $n,m\to\infty$. We then provide an explicit computation for some more general mixed-moments by considering the position of matrix $\X$ in the product $\X^{\eps_1}\cdots\X^{\eps_k}$. We, therefore, deduce closed-form formulas for some mixed-moments of Ginibre matrices. 
\end{abstract}
\maketitle

\section{Introduction}

\subsection{General setting}

In the past few decades, the study of large random matrices has played an increasingly central role in a growing number of fields. In deep learning, the large number of parameters needed to model the different layers of linear neural networks are packaged into large matrices, which one can approximate by random matrices to some extent \cite{saxe_exact_2014}. Furthermore, the study of the statistical features of financial time-series $(x^i_t)_{1\leq t\leq T,1\leq i\leq N}$ requires the understanding of its sample covariance matrix $\mathbf{E}=T^{-1}\mathbf{H}\mathbf{H}^\top$. The matrix $\mathbf{H}$ encapsulates the different observations of the time-series $H_{it}=x^i_t$ which can be modeled by random variables (see for instance \cite{potters_bouchaud_2020}). Another example is the celebrated study by Robert May \cite{may1972will} in the fields of complex systems and theoretical ecology. Studying the stability of large linear systems $\d\x/\d t=\mathbf{A}\x$ where $\A$ is an $N\times N$ random matrix with Gaussian entries, he showed that the stability of the system only depends on $N$ and $\sigma^2=\E{A_{ij}^2}$ (see \cite{Fyodorov_Khoruzhenko_2016} for recent developments of the stability of random \emph{non-linear} systems, and \cite{Bizeul2020PositiveSF, Biroli_2018} for a connection with ecological equilibria). More than a mere tool, studying random matrices has grown into an independent field of research known as Random Matrix Theory (RMT).

At the origin of RMT, the physicist Eugene Wigner used random matrices to model the Hamiltonian of heavy nuclei. Although deterministic, these Hamiltonian were so complex that Wigner had the novel idea to replace them with specific elements of an ensemble of random matrices, the so-called Wigner matrices. A Wigner matrix $\W$ is a $N\times N$ symmetric random matrix whose elements $\{\W_{ij}\}_{1\leq i<j\leq N}$ and $\{\W_{ii}\}_{1\leq i\leq N}$ are complex-valued or real-valued \iid~random variables with zero mean and finite second moment. These matrices have been extensively studied (see for instance \cite{anderson_introduction_2009} for a review), and one of their most striking features is that they display a universal behavior in the large dimensional limit $N\to\infty$. Calling $\{\lambda^N_i\}_{1\leq i\leq N}$ the eigenvalues of $\W/\sqrt{N}$, the spectral measure $\mu^N_{\W/\sqrt{N}}(\lambda)=N^{-1}\sum_{i=1}^N\delta(\lambda-\lambda_i^N)$  converges to the so-called \emph{semi-circle} distribution
\begin{equation}
   \mu^N_{\W/\sqrt{N}}(\lambda)\underset{N\to\infty}{\longrightarrow} \mu_{\textsc{s.c}}(\lambda)=\frac{1}{2\pi}\sqrt{4-\lambda^2}\mathds{1}(\lambda^2\leq 2),
\end{equation}
regardless of the details of the distribution of the elements $\W_{ij}$. This result, due to Wigner himself \cite{Wigner1955CharacteristicVO,wigner1967random}, showed that \emph{universality} looms underneath the apparent complexity and disparity of random matrices, and sparked a vivid interest in these objects. 

The Wigner semi-circle distribution is a particular example of a limiting distribution for the eigenvalues of large random matrices. Other such distributions $\mu$ exist (Wishart \cite{wishart_generalised_1928}, circular \cite{ginibre_statistical_1965,sommers_asymmetric}, Jacobi \cite{ramli_spectral_2012} to name a few), each one corresponding to a specific ensemble of random matrices $\A$. Each of them is characterized by its moments $m_k$ defined as
\begin{equation}
    m_k=\lim_{N\to\infty}\frac{1}{N}\E{\Tr{\A^k}}=\int\d\lambda\mu(\lambda)\lambda^k,
\end{equation}
where $\E{\cdot}$ is the expectation over all matrices $\A$. Remarkably, in the large dimensional limit, the normalized trace $\tr{\cdot}:=N^{-1}\Tr{\cdot}$ does not fluctuate from samples to samples $\A$, and, for large enough matrices $\A$, $m_k\approx\frac{1}{N}\Tr{\A^k}$.

In the present article, we focus on more general quantities called mixed-moments which consist in computing of the normalized trace of a product of random matrices, i.e.
\begin{equation}
    \varphi(\A_1,\ldots,\A_k)=\E{\tr{\A_1\cdots\A_k}}.
    \label{eq:mixed_moments_general}
\end{equation}
These quantities play a central role in the statistical physics of disordered systems. For instance, considering the one-dimensional Ising model over $N$ spins on a chain with random couplings $J_i$ in a random magnetic field $h_i$ with Hamiltonian
\[\textlmss{H}=-\sum_{i}J_{i}\sigma_{i+1}\sigma_i+h_i\sigma_i,\]
one can show that the partition function $\Zc_N(\beta)=\sum_{\{\sigma\}}e^{-\beta\textlmss{H}}$ can be written as a mixed-moment of random transfer matrices $\textlmss{L}_i$
\begin{equation}
    \Zc_N(\beta)=\Tr{\prod_i\textlmss{L}_i},\quad\textlmss{L}_i=\begin{pmatrix}
e^{\beta(J_i+h_i)} & e^{\beta(J_i-h_i)}\\
e^{-\beta(J_i+h_i)} & e^{-\beta(J_i-h_i)}
\end{pmatrix}.
\end{equation}
See \cite{crisanti2012products} for an extensive review of concerning products of random matrices in statistical physics. More recently, in \cite{Dessertaine2022Cones}, products of random matrices have arisen again in the context of cone-wise linear systems. Consider two matrices $\A$ and $\B$ such that a vector $\mathbf{v}\in\R^N$ evolves according to 
\begin{equation}
    \mathbf{v}(t+1)=\begin{cases}
    \A\mathbf{v}(t)&, v_1(t)>0\\
    \B\mathbf{v}(t)&, v_1(t)<0
    \end{cases}.
\end{equation}
Whenever $\A$ and $\B$ are orthogonally invariant\footnote{A matrix $\X$ is said to be orthogonally invariant if $\O\X\O^\top\overset{d}{=}\X$ for any orthogonal matrix $\O$, where $\overset{d}{=}$ denotes the equality in distributions.}, in either region $v_1(t)>0$ or $v_1(t)<0$, the vector $\mathbf{v}$ can be mapped onto a centered Gaussian process with covariance
\begin{equation}
    \langle v_i(t)v_j(s) \rangle \underset{N\to\infty}{\longrightarrow}\delta_{ij}\tr{\X^t(\X^\top)^s},
    \label{eq:conewise}
\end{equation}
with $\X=\A,\B$ depending on the region. The spectral density of $\A$ or $\B$ determines the (random) time spent in either region, also called persistence time (see \cite{BrayReviewPersistence} for an extensive review in the context of out-of-equilibrium systems), which translates, for a large class of random matrices, into the non-self-averaging of the maximal Lyapunov of the system.

For any products, \eqref{eq:mixed_moments_general} can be expressed through the moment-cumulant formula (see below, and \cite{nica_speicher_2006, potters_bouchaud_2020}), which, in general, is quite an involved computation. It can be simplified if matrices $\A_\ell$ are \emph{free}, i.e. randomly rotated from one another, as in the spin chain example but remains complex for cone-wise linear systems \eqref{eq:conewise}, since $\X$ and $\X^\top$ are not independent.

\subsection{Gaussian Elliptic Matrices}

In this work, we are interested in computing the mixed-moments of a matrix $\X$ drawn from the Gaussian Elliptic Ensemble (\gee). Let $\W$ be a $N\times N$ random matrix drawn from the Gaussian Orthogonal Ensemble (\goe), i.e. a symmetric rotationally invariant matrix whose elements $\W_{ij}$ are Gaussian random variables verifying:
\begin{equation}
\mathbb{E}\left[\W_{ij}\right]=0,\quad\mathbb{E}\left[\W_{ii}^2\right]=\frac{2\sigma^2}{N},\quad\mathbb{E}\left[\W_{ij}^2\right]=\frac{\sigma^2}{N}.
    \label{eq:def_goe}
\end{equation}
Without loss of generality, we will set $\sigma=1$ in the following. Note that \goe is a particular case of Wigner matrices. Taking $\W_1$ and $\W_2$ to be two \emph{free} \goe~matrices, a \gee~matrix $\X$ of parameter $\rho$ is defined as 
\begin{equation}
    \X=\sqrt{\frac{1+\rho}{2}}\W_1+i\sqrt{\frac{1-\rho}{2}}\W_2.
\end{equation}
The entries of $\X$ are complex-valued Gaussian random variables that satisfy:
\begin{equation}
    \mathbb{E}\left[\X_{ij}\right]=0,\quad\mathbb{E}\left[|\X|_{ii}^2\right]=\frac{1+\rho}{N},\quad\mathbb{E}\left[\X_{ij}\X^*_{ji}\right]=\frac{\rho}{N}.
    \label{eq:prop_el_el}
\end{equation}
The elliptic parameter $\rho$ interpolates between three different ensembles of random matrices: $(a)$ hermitian ($\X^\dagger=\X$) random matrices from \goe~when $\rho=1$, $(b)$ anti-hermitian ($\X^\dagger=-\X)$ random matrices from anti-\goe~when $\rho=-1$, and $(c)$ random matrices with uncorrelated entries $\X_{ij}$ and $\X^*_{ji}$ from the complex Ginibre ensemble when $\rho=0$. Elliptic matrices have been introduced as the first study of non-Hermitian RMT \cite{sommers_asymmetric}. The authors show that the spectral density of \gee~ converges towards an ellipse centered at zero and spanning $1+\rho$ over the real axis and $1-\rho$ over the imaginary axis. These matrices have been especially studied in the context of theoretical ecology to model interaction matrices between competing species, see \cite{Biroli_2018,bunin2016interaction}.

For $k$-tuples of integers $\n_k=(n_1,\ldots,n_k)$ and $\m_k=(m_1,\ldots,m_k)$~\footnote{We will drop the index $k$ on $\n_k$ and $\m_k$ when there is no ambiguity.}, we denote by $\mm{\n}{\m}(\rho)$ the mixed-moment
\begin{equation}
    \mm{\n}{\m}(\rho)=\varphi\left(\X^{n_1},(\Xt)^{m_1},\cdots,\X^{n_k},(\Xt)^{m_k}\right).
    \label{eq:mm_general}
\end{equation}
Whenever $\rho=1$, i.e. $\X=\W_1$ is drawn from \goe, the previous formula reduces to the moment of $\W_1$ of order $|\n|+|\m|:=\sum_{i}n_i+\sum_{i}m_i$ which are given by
\begin{equation}
    \mm{\n}{\m}(1)=\varphi\left(\X^{|\n|+|\m|}\right)=\begin{dcases}
        C_{(|\n|+|\m|)/2} &,\; |\n|+|\m|\text{ even}\\
        0 &,\; |\n|+|\m|\text{ odd}
    \end{dcases},
\end{equation}
where $C_n=\frac{1}{n+1}\binom{2n}{n}$ is the $n$-th Catalan number.

In the particular case where $\X$ is a Ginibre matrix ($\rho=0$), the computation of $\mm{\n}{\m}(0)$ should yield the solution the enumeration of non-crossing bit-string pairings \cite{kemp_enumeration_2011,schumacher_enumeration_2013}, otherwise known as the Knights and Ladies of the Round Table: considering two groups of people ($1$s and $0$s) seated around a table, how many ways are there to pair them up in conversations so that no two conversations cross. Perhaps surprisingly, a closed-form solution has yet to be found for this problem, although some specific cases are known \cite{kemp_enumeration_2011}, for instance when $\n_k=\m_k=(r,\ldots,r)$, one gets
\begin{equation}
    \mm{(r,\ldots,r)}{(r,\ldots,r)}(0)=FC_{r}(k),
\end{equation}
with $FC_{r}(k)$ a Fuss-Catalan number. In recent work \cite{halmagyi_mixed_2020}, authors have focused on the study of matrices $\X_{(n)}$ given by the product of $n$ Ginibre matrices $\X_{(n)}=A_1\cdots A_n$. They found that the mixed-moments of such matrices are given by the enumeration of non-crossing pairings weighted by Fuss-Catalan numbers \cite{Mlotkowski2010}. As highlighted by the authors, considering \gee~rather than the Ginibre ensemble relaxes the stiff \iid~condition of the Ginibre ensemble and might allow to carry out the computation. We hope the ideas put forward in this paper can contribute to this end.

\subsection{Outline of the paper and main results}

This paper is organized as follows. In section \secref{sec:sec1}, we establish a general formula for $\mm{n}{m}(\rho)$ using the moment-cumulant formula \cite{nica_speicher_2006,potters_bouchaud_2020}. In \secref{sec:sec2}, we focus on a specific class of mixed-moments $\mm{n}{m}(\rho)=\varphi\left(\X^{n},(\Xt)^{m}\right)$. Using two different approaches, we prove the first main result of this paper by expressing these mixed-moments as polynomials in $\rho$ whose coefficients are integers known as Catalan triangular numbers \cite{carlitz1972sequences}:
\begin{equation}
    \mm{n}{m}(\rho)=\begin{dcases*}
    \;\sum_{k=0}^{u\wedge v}\rho^{u+v-2k}\frac{(2k+1)^2}{(u+k+1)(v+k+1)}\binom{2u}{u+k}\binom{2v}{v+k} & \text{if $n=2u,m=2v$}\\
    \;\sum_{k=0}^{u\wedge v}\rho^{u+v-2k}\frac{4(k+1)^2}{(u+k+2)(v+k+2)}\binom{2u+1}{u+k+1}\binom{2v+1}{v+k+1}& \text{if $n=2u+1,m=2v+1$}\\
    \;0& \text{otherwise},
    \end{dcases*}
\end{equation}
where $a\wedge b=\min(a,b)$. Then, in \secref{sec:sec3}, we study the asymptotic behavior of the of $\mm{n}{m}(\rho)$ as $n,m\to\infty$. Finally, we study in \secref{sec:sec4} more general classes of mixed-moments. We establish several formulas, consisting in the second class of main results, by considering the position of matrices $\X$ in the product $\X^{n_1}(\Xt)^{m_1}\cdots\X^{n_k}(\Xt)^{m_k}$. More specifically, we encode these positions in a tuple $\i$ and we establish formulas when the number of even elements in $\i$ is at most $2$. These specific cases, which are analytically tractable, also provide the mixed-moments of Ginibre matrices when there are as many matrices $\X$ and $\Xt$.

\section{General formula for the mixed-moments\label{sec:sec1}}

\subsection{An enumerative formula}

Considering random matrices $\A_1,\ldots,\A_N$, it is well known \cite{nica_speicher_2006, potters_bouchaud_2020} that the mixed-moment $\varphi{\A_1,\cdots,\A_k}$ is related to the mixed-cumulants $\kappa_\ell{\left[\A_1\cdots\A_k\right]}$ via the moment-cumulant formula
\begin{equation}
    \varphi(\A_1,\cdots,\A_k)=\sum_{\pi\in{\nc}(k)}\prod_{V\in\pi}\kappa_\pi(V)[\A_1,\cdots,\A_k].
    \label{eq:moment-cumulant}
\end{equation}
The sum runs over all non-crossing partitions $\pi\in\nc (k)$ of $[\![ 1, k]\!]$, i.e. all possible allocations of the digits between $1$ and $k$ into subgroups $V$, called blocks, such that there would be no crossings if one were to draw some edges linking the elements belonging to the same group. The quantity $\kappa_\pi(V)[\A_1,\cdots,\A_k]$ denotes the mixed-cumulant of matrices $\{\A_i\}$ which are linked within block $V$ of $\pi$. For instance if $k=4$, $\pi=\{(1,2,4),(3)\}$ is a valid non-crossing partition with two blocks $V_1=(1,2,4)$, $V_2=(3)$, and associated cumulants $\kappa_3[\A_1,\A_2,\A_4]$ and $\kappa_1[\A_3]$. More examples can be found on \figref{fig:non-crossing-partition}.

Let us focus on the case where $\A_i=\X^{\epsilon_i}$ where $\epsilon_i\in\{1,\dagger\}$ (in the sense that $\X^{\epsilon_i}$ is either $\X$ or $\X^\dagger$). In the definition $\X=\sqrt{(1+\rho)/2}\W_1+i\sqrt{(1-\rho)/2}\W_2$, $\W_1$ and $\W_2$ are assumed mutually free which implies that any mixed-cumulant between $\W_1$ and $\W_2$ vanish \cite{nica_speicher_2006}, e.g. $\kappa_3[\W_1,\W_1,\W_2]=\kappa_2[\W_1,\W_2,\W_2]=0$. Consequently, considering $(\eps_i,\ldots,\eps_\ell)\in\{1,\dagger\}^\ell$ such that $\sum_j \mathds{1}(\eps_j=\dagger)=r$, we have
\begin{equation}
\kappa_\ell\left[\X^{\eps_{i_1}},\ldots,\X^{\eps_{i_\ell}}\right]=\left(\frac{1+\rho}{2}\right)^{\ell/2}\kappa_\ell[\W_1]+i^{\ell}(-1)^{r}\left(\frac{1-\rho}{2}\right)^{\ell/2}\kappa_{\ell}[\W_2],
    \label{eq:mc_gem1}
\end{equation}
where, conventionally, $\kappa_{\ell}[\A]:=\kappa_{\ell}[\A,\ldots,\A]$, where $\A$ is repeated $\ell$ times. For a \goe~matrix $\W$, one can show that (see \cite{nica_speicher_2006})
\begin{equation}
    \kappa_\ell[\W]=\begin{cases}
        1 &, \;\ell=2\\
        0 &, \;\ell\neq 2
    \end{cases},
\end{equation}
such that \eqref{eq:mc_gem1} vanishes whenever $\ell\neq 2$. On the other hand, plugging $\ell=2$ in \eqref{eq:mc_gem1} yields
\begin{equation}
\kappa_2\left[\X^{\eps_{i_1}},\X^{\eps_{i_2}}\right]=\begin{cases}
        1 &, \;\eps_1\neq\eps_2\\
        \rho &, \;\eps_2=\eps_2
    \end{cases}.
    \label{eq:cumulant_circular}
\end{equation}
As a consequence, in the present case where $\A_i=\X^{\eps_i}$, the sum from \eqref{eq:moment-cumulant} actually runs over all non-crossing \emph{pairings} over $[\![ 1, k]\!]$, i.e. non-crossing partitions for which every block is a pair $(r,s)$. Using \eqref{eq:cumulant_circular}, we finally find
\begin{equation}
    \tau\left(\X^{\epsilon_1}\cdots\X^{\epsilon_k}\right)=\sum_{\pi\in\nc_2(k)}\rho^{\sigma(\pi)},
    \label{eq:final_formula}
\end{equation}
where $\nc_2(k)$ denotes the set of all non-crossing pairings over $[\![ 1, k]\!]$ and $\sigma(\pi)=\card{\{(r,s)\in\pi,\;\epsilon_r=\epsilon_r\}}$\footnote{$\card{A}$ denotes the cardinality of set $A$.}. This result can be found in \cite[Lemma~1]{adhikari2019brown}. 

\begin{figure}[t]
    \centering
    \begin{tikzpicture}
    \draw (0,0) -- (0,1) -- (1,1) -- (1,0);
    \draw (2,0) -- (2, 1) -- (3,1) -- (3,0);
    \node[fill=white, opacity=1] at (0,0) {$\A_1$};
    \node[fill=white, opacity=1] at (1,0) {$\A_2$};
    \node[fill=white, opacity=1] at (2,0) {$\A_3$};
    \node[fill=white, opacity=1] at (3,0) {$\A_4$};
    \node at (1.5,-1) {$\pi=\{(1,2),(3,4)\}$};
    \end{tikzpicture}\quad
    \begin{tikzpicture}
    \draw (0,0) -- (0,1) -- (3,1) -- (3,0);
    \draw (1,0) -- (1,1);
    \draw (2,0) -- (2, 0.75);
    \node[fill=white, opacity=1] at (0,0) {$\A_1$};
    \node[fill=white, opacity=1] at (1,0) {$\A_2$};
    \node[fill=white, opacity=1] at (2,0) {$\A_3$};
    \node[fill=white, opacity=1] at (3,0) {$\A_4$};
    \node at (1.5,-1) {$\pi=\{(1,2,4),(3)\}$};
    \end{tikzpicture}\quad
    \begin{tikzpicture}
    \draw (0,0) -- (0,1) -- (3,1) -- (3,0);
    \draw (1,0) -- (1,1);
    \draw (2,0) -- (2, 1);
    \node[fill=white, opacity=1] at (0,0) {$\A_1$};
    \node[fill=white, opacity=1] at (1,0) {$\A_2$};
    \node[fill=white, opacity=1] at (2,0) {$\A_3$};
    \node[fill=white, opacity=1] at (3,0) {$\A_4$};
    \node at (1.5,-1) {$\pi=\{(1,2,3,4)\}$};
    \end{tikzpicture}
    \quad
    \begin{tikzpicture}
    \draw (0,0) -- (0,1) -- (2,1) -- (2,0);
    \draw (1,0) -- (1,1.1) -- (3,1.1) -- (3,0);
    \node[fill=white, opacity=1] at (0,0) {$\A_1$};
    \node[fill=white, opacity=1] at (1,0) {$\A_2$};
    \node[fill=white, opacity=1] at (2,0) {$\A_3$};
    \node[fill=white, opacity=1] at (3,0) {$\A_4$};
    \node at (1.5,-1) {$\pi=\{(1,3),(2,4)\}$};
    \end{tikzpicture}
    \begin{tikzpicture}[scale=0.6, every node/.style={scale=1}]
    \draw (5,-5) to[bend left=60] (10,-5);
    \draw (6,-5) to[bend left=60] (7,-5);
    \draw (8,-5) to[bend left=60] (9,-5);
    \draw (5,-5)  node[ptm={white}] {};
    \draw (6,-5) node[ptm={black}] {};
    \draw (7,-5) node[ptm={white}] {};
    \draw (8,-5) node[ptm={white}] {};
    \draw (9,-5) node[ptm={black}] {};
    \draw (10,-5) node[ptm={black}] {};
    \node at (11.5,-5) {$\Longleftrightarrow$};
    \CircleMatching{6}{1/6/white/black,2/3/black/white/,4/5/white/black}{1}{14,-5};
    \begin{scope}[xshift=4cm]
        \draw (14,-5) to[bend left=60] (15,-5);
        \draw (16,-5) to[bend left=60] (17,-5);
        \draw (18,-5) to[bend left=60] (19,-5);
        \draw (14,-5)  node[ptm={black}] {};
        \draw (15,-5) node[ptm={white}] {};
        \draw (16,-5) node[ptm={black}] {};
        \draw (17,-5) node[ptm={white}] {};
        \draw (18,-5) node[ptm={white}] {};
        \draw (19,-5) node[ptm={black}] {};
    \end{scope}
    \node at (16.5,-5) {$\Longleftrightarrow$};
    \end{tikzpicture}
    \caption{Top: four examples of partitions over 4 elements. From left to right, the different contributions to $\varphi(\A_1,\A_2,\A_3,\A_4)$ are: $\kappa_2(\A_1,\A_2)\kappa_2(\A_3,\A_4)$, $\kappa_3(\A_1,\A_2,\A_4)\kappa_1(\A_3)$, $\kappa_4(\A_1,\A_2,\A_3,\A_4)$ and 0. The first three are non-crossing and the rightmost is crossing. Bottom: equivalence between two line representations after a cyclic permutation in the computation of $\tr{\X\Xt\X^2(\Xt)^2}$. The circle representation lifts the ambiguity.}
    \label{fig:non-crossing-partition}
\end{figure}
In the following, we will represent non-crossing partitions diagrammatically by drawing some edges between points spread over a line or a circle. Each point will bear a color encoding the corresponding matrix: $\circ$ for matrices $\X$, and $\bullet$ for matrices $\Xt$. Due to the invariance under cyclic permutations of $\tr{\cdot}$, we will privilege the circle representation in the following, see \figref{fig:non-crossing-partition}. The first fact to notice is that whenever $n$ is odd the set $\nc_2(n)$ is empty since it is not possible to construct pairings across an odd number of elements. Consequently, the mixed-moments involving an odd number of elliptic terms $\X^{\epsilon_i}$ are always zero. Unless stated otherwise, we will implicitly consider an even number of terms, which corresponds to $|\n|$ and $|\m|$ being either both even, or both odd.

\subsection{From exponential to polynomial complexity}

As an example of the application of \eqref{eq:final_formula}, let us consider the computation of the mixed-moment $\mm{(2,1,1)}{(1,1,2)}(\rho)=\tau{\left(\X^2\Xt\X\Xt\X(\Xt)^2\right)}$. There are $8$ elliptic terms involved which corresponds to a sum over the $C_4=14$ parings of $NC_2(8)$ in \eqref{eq:final_formula} such that
\begin{equation}
    \begin{aligned}
\tikz[scale=0.6, every node/.style={scale=1}]{\node at (0,0) {};
\node at (0.5,0.95) {$\mm{(2,1,1)}{(1,1,2)}(\rho)$}}&
\tikz[scale=0.6, every node/.style={scale=1}]{
\node at (-1.65,0) {=};
\CircleMatching{8}{1/8/white/black,2/7/white/black,3/6/black/white,4/5/white/black}{1}{0,0};}
\;
\tikz[scale=0.6, every node/.style={scale=1}]{
\node at (-1.65,0) {+};
\CircleMatching{8}{1/8/white/black,2/7/white/black,3/4/black/white,5/6/black/white}{1}{0,0};}\;
\tikz[scale=0.6, every node/.style={scale=1}]{
\node at (-1.65,0) {+};
\CircleMatching{8}{1/8/white/black,2/3/white/black,4/7/white/black,5/6/black/white}{1}{0,0};}
\;
\tikz[scale=0.6, every node/.style={scale=1}]{
\node at (-1.65,0) {+};
\CircleMatching{8}{1/8/white/black,2/3/white/black,4/5/white/black,6/7/white/black}{1}{0,0};}
\;
\tikz[scale=0.6, every node/.style={scale=1}]{
\node at (-1.65,0) {+};
\CircleMatching{8}{1/8/white/black,2/5/white/black,3/4/black/white,6/7/white/black}{1}{0,0};}
\;
\tikz[scale=0.6, every node/.style={scale=1}]{
\node at (-1.65,0) {+};
\CircleMatching{8}{1/2/white/white,3/8/black/black,4/7/white/black,5/6/black/white}{1}{0,0};}
\;
\tikz[scale=0.6, every node/.style={scale=1}]{
\node at (-1.65,0) {+};
\CircleMatching{8}{1/2/white/white,3/8/black/black,4/5/white/black,6/7/white/black}{1}{0,0};}\\
&
\tikz[scale=0.6, every node/.style={scale=1}]{
\node at (-1.65,0) {\phantom{=}};
\CircleMatching{8}{1/2/white/white,3/4/black/white,5/8/black/black,6/7/white/black}{1}{0,0};}
\;
\tikz[scale=0.6, every node/.style={scale=1}]{
\node at (-1.65,0) {+};
\CircleMatching{8}{1/4/white/white,2/3/white/black,5/8/black/black,6/7/white/black}{1}{0,0};}
\;
\tikz[scale=0.6, every node/.style={scale=1}]{
\node at (-1.65,0) {+};
\CircleMatching{8}{1/6/white/white,2/5/white/black,3/4/black/white,7/8/black/black}{1}{0,0};}
\;
\tikz[scale=0.6, every node/.style={scale=1}]{
\node at (-1.65,0) {+};
\CircleMatching{8}{1/6/white/white,2/3/white/black,4/5/white/black,7/8/black/black}{1}{0,0};}
\;
\tikz[scale=0.6, every node/.style={scale=1}]{
\node at (-1.65,0) {+};
\CircleMatching{8}{1/2/white/white,3/6/black/white,4/5/white/black,7/8/black/black}{1}{0,0};}
\;
\tikz[scale=0.6, every node/.style={scale=1}]{
\node at (-1.65,0) {+};
\CircleMatching{8}{1/4/white/white,2/3/white/black,5/6/black/white,7/8/black/black}{1}{0,0};}
\;
\tikz[scale=0.6, every node/.style={scale=1}]{
\node at (-1.65,0) {+};
\CircleMatching{8}{1/2/white/white,3/4/black/white,5/6/black/white,7/8/black/black}{1}{0,0};}\\
&
\tikz[scale=0.6, every node/.style={scale=1}]{
\node (0,0) {$5+9\rho^2$.};
\node at (-1.65,0) {=};
}
\end{aligned}
\label{eq:example_p_(211)_(112)}
\end{equation}

Even though this formula works fine, the number of terms involved in the summation in \eqref{eq:final_formula}, given by the $k/2$-th Catalan number, grows as $\Oc\left(4^{k/2}k^{-3/2}\right)$ which quickly becomes impossible to handle, even numerically. However, in example \eqref{eq:example_p_(211)_(112)}, we can see that a lot of the non-crossing pairings actually have the same contributions to \eqref{eq:final_formula}. In general, the sum can be expressed as a polynomial in $\rho$ whose coefficients can be expressed by studying $\sigma(\cdot)$.

In the following, we will denote by $\nc_2(\n,\m)$ the set of non-crossing pairings between $|\n|+|\m|$ elements, where each element bears a color ($\circ$ or $\bullet$). The coloring follows the specification from $\n$ and $\m$, i.e. $n_1$ $\circ$ followed by $m_1$ $\bullet$ then $n_2$ $\circ$ followed by $m_2$ $\bullet$ etc. For instance $\nc_2((1,2),(1,3))$ is the set of non-crossing pairings over $(\circ\,\bullet\,\bullet\,\circ\,\bullet\,\bullet\,\bullet)$. Now, in anticipation of the mapping onto Temperley-Lieb diagrams in \secref{sec:sec2}, it makes more sense to study the complementary function to $\sigma$, which we denote by $\sigma_c$ and counts the number of pairs $(\circ,\bullet)$ between elements of different colors in a pairing $\pi$, i.e.
\[\sigma_c(\pi)=\card{\{(r,s)\in\pi,\;\epsilon_r\neq\epsilon_r\}}.\]
We list here some properties of $\sigma$ and $\sigma_c$:
\begin{enumerate}[label=$(\roman*)$]
    \item The two functions are linked through the relation $\sigma+\sigma_c=(|\n|+|\m|)/2:=L$, where $L\in\N$ is the total number of pairs for any $\pi\in\nc_2(\n,\m)$.
    \item $\sigma_c$ is always of the same parity as $|\n|$ and $|\m|$. Indeed, assume, for exposition, that $|\n|,|\m|\in 2\N$ and there exists $\pi\in\nc_2(\n,\m)$ such that $\sigma_c(\pi)=2k+1$ for some $k$. The number of pairs $(\circ,\bullet)$ being odd, there are $|\n|-2k-1\in 2\N+1$ $\circ$-elements and $|\m|-2k-1\in 2\N+1$ $\bullet$-elements remaining to be paired. However, the remaining elements must be paired with an element of the same color (since otherwise $\sigma_c(\pi)>2k+1$), which is impossible since both $|\n|-2k-1$ and $|\m|-2k-1$ are odd.
    \item$\sigma_c$ is smaller than $K:=\min(|\n|,|\m|)$. Indeed, assuming $K=|\m|$, as soon as all $K$ $\bullet$-elements are paired with $K$ $\circ$-elements, the remaining $|\n|-K\geq0$ $\circ$-elements have to be paired among themselves, such that $K$ is the maximum of $\sigma_c$.
\end{enumerate}
It is now easy to see that furnish a natural partitioning of $\nc_2(\n,\m)$ given by:
\begin{subequations}
\begin{align}
     \nc_2(\n,\m)&=
    \bigsqcup_{k=0}^{K/2}\nc^{2k}_2(\n,\m),\quad |\n|, |\m|\in2\N\label{eq:even_partition}\\
    \nc_2(\n,\m)&=\bigsqcup_{k=0}^{(K-1)/2}\nc^{2k+1}_2(\n,\m),\quad |\n|, |\m|\in2\N+1,
\end{align}
\end{subequations}
where we introduced the sets $\nc^{\ell}_2(\n,\m)\subseteq\nc_2(\n,\m)$ over which $\sigma_c$ assumes a constant value $\ell$, i.e.
\begin{equation}
    \nc^{\ell}_2(\n,\m)=\left\{\pi\in\nc_2(\n,\m),\;\sigma_c(\pi)=\ell\right\},
\end{equation}
and where $\bigsqcup$ denotes the disjoint union. For simplicity, consider the case where $|\n|, |\m|\in2\N$ in the following. Using \eqref{eq:even_partition} and properties $(i)-(iii)$, we can rework \eqref{eq:final_formula} as
\begin{align*}
    \mm{\n}{\m}(\rho)=\sum_{\pi\in\nc_2(\n,\m)}\rho^{\sigma(\pi)}=\sum_{\pi\in\nc_2(\n,\m)}\rho^{L-\sigma_c(\pi)}=\sum_{k=0}^{K/2}\sum_{\pi\in\nc^{2k}_2(\n,\m)}\rho^{L-\sigma_c(\pi)},
\end{align*}
such that
\begin{equation}
    \mm{\n}{\m}(\rho)=\sum_{k=0}^{K/2}\rho^{L-2k}\card{\nc^{2k}_2(\n,\m)}.
    \label{eq:poly_formula}
\end{equation}
As stated above, this formula reduces the complexity of computing  \eqref{eq:final_formula} from exponential to polynomial which is a substantial improvement. However, it requires knowing the cardinality of the sets $\nc^{2k}_2(\n,\m)$, which \emph{a priori} is not straightforward. For instance, in example \eqref{eq:example_p_(211)_(112)}, we can see that
\[\card{\nc^{0}_2((2,1,1),(1,1,2))}=5\;,\;\card{\nc^{2}_2((2,1,1),(1,1,2))}=9\;,\;\card{\nc^{4}_2((2,1,1),(1,1,2))}=0.\]
However, a simple permutation $(2,1,1)\to(1,2,1)$ in $\n$ would yield
\[\card{\nc^{0}_2((1,2,1),(1,1,2))}=4\;,\;\card{\nc^{2}_2((1,2,1),(1,1,2))}=9\;,\;\card{\nc^{4}_2((1,2,1),(1,1,2))}=1.\]
In the following, we will focus on the case where $\n=(n)$ and $\m=(m)$ and provide an explicit formula for the cardinality of $\nc^{\ell}_2(n,m)$.

\section{\texorpdfstring{Explicit formula for $\mm{n}{m}(\rho)$}{}\label{sec:sec2}}

Let us consider the case $k=1$ in \eqref{eq:mm_general}, i.e. the specific case in \eqref{eq:final_formula} where $\epsilon_i=1$ for $1\leq i\leq n_1$ and $\epsilon_i=\dagger$  for $n_1+1\leq i\leq n_1+m_1$. We will denote by $n,m$ the exponent $n_1,m_1$ in \eqref{eq:mm_general}. For clarity, we will only consider the case where $n,m\in 2\N$ since the odd case can be treated with the same methods. We set $n=2u$, $m=2v$, and assume (without loss of generality using the invariance of $\tr{\cdot}$ under trans-conjugation) $\min(n,m)=m$. In this section, we will show that the sets $\nc_2^\ell(n,m)$ have a cardinality given by Catalan triangular numbers (see Appendix \ref{ap:catalan_triangle}) such that
\begin{equation}
    \card{\nc_2^{2\ell}(2u,2v)}=\frac{(2k+1)^2}{(u+k+1)(v+k+1)}\binom{2u}{u+k}\binom{2v}{v+k}.
    \label{eq:objective_combinatorics}
\end{equation}
In the following, we will order and label the elements $\circ,\bullet$ to pair such that
\begin{equation*}
    \begin{tikzpicture}
    \node[ptm={white}, label=below:$1$] at (0,0) {};
    \node[ptm={white}, label=below:$2$] at (-0.5,0) {};
    \node[ptm={white}, label=below:$n$] at (-4,0) {};
    \node[] at (-2.25,0) {$\ldots$};
    \node[ptm={black}, label=below:$1$] at (0.5,0) {};
    \node[ptm={black}, label=below:$2$] at (1,0) {};
    \node[ptm={black}, label=below:$m$] at (3,0) {};
    \node[] at (2,0) {$\ldots$};
    \node[] at (4,0) {$\Longleftrightarrow$};
    \draw (6.5,0) pic{carc=-360*8/20:360*0/20:1.5cm};
    \draw (6.5,0) pic{carc=360*2/20:360*10/20:1.5cm};
    \node[ptm={white}, label=below:$1$] at ($(360*16/20:1.5)+(6.5,0)$) {};
    \node[ptm={black}, label=below:$1$] at ($(360*15/20:1.5)+(6.5,0)$) {};
    \node[ptm={white}, label=below:$2$] at ($(360*17/20:1.5)+(6.5,0)$) {};
    \node[ptm={black}, label=below:$2$] at ($(360*14/20:1.5)+(6.5,0)$) {};
    \node[ptm={white}, label=above:$n$] at ($(360*7/20:1.5)+(6.5,0)$) {};
    \node[ptm={black}, label=above:$m$] at ($(360*8/20:1.5)+(6.5,0)$) {};
    
    \node[circle, draw=black, fill=black, inner sep=0pt, outer sep=0pt, minimum size=0.3mm] at ($(360*1/20:1.5)+(6.5,0)$) {};
    \node[circle, draw=black, fill=black, inner sep=0pt, outer sep=0pt, minimum size=0.3mm] at ($(360*0.5/20:1.5)+(6.5,0)$) {};
    \node[circle, draw=black, fill=black, inner sep=0pt, outer sep=0pt, minimum size=0.3mm] at ($(360*1.5/20:1.5)+(6.5,0)$) {};
    \node[circle, draw=black, fill=black, inner sep=0pt, outer sep=0pt, minimum size=0.3mm] at ($(360*11/20:1.5)+(6.5,0)$) {};
    \node[circle, draw=black, fill=black, inner sep=0pt, outer sep=0pt, minimum size=0.3mm] at ($(360*11.5/20:1.5)+(6.5,0)$) {};
    \node[circle, draw=black, fill=black, inner sep=0pt, outer sep=0pt, minimum size=0.3mm] at ($(360*10.5/20:1.5)+(6.5,0)$) {};
    \node[] at (8.5,0) {.};
    \end{tikzpicture}
\end{equation*}

\subsection{\texorpdfstring{Cardinality of $\nc_2^\ell(n,m)$: a combinatorial approach}{}}

\begin{figure}
    \centering
    \begin{tikzpicture}[every node/.style={scale=1}]
    \include{combinatorial_proof}
    \node[above] at (current bounding box.north east) {$(a)$};
    \end{tikzpicture}
    \quad
    \begin{tikzpicture}[every node/.style={scale=1}]
    \input{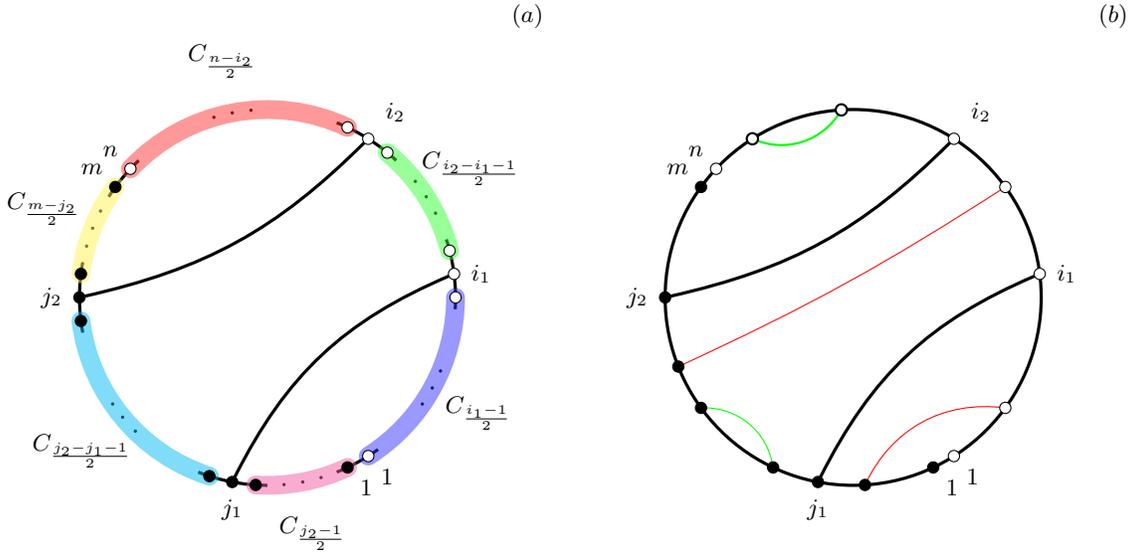}
    \node[above] at (current bounding box.north east) {$(b)$};
    \end{tikzpicture}
    \caption{Example of the enumeration of the pairings in $\nc_2^{2}(n,m)$ with given mixed pairs $(i_1,j_1)$ and $(i_2,j_2)$. $(a)$ Natural splitting of the remaining elements needing to be paired. Regions dark-blue and magenta correspond to $\nu_0$ and $\mu_0$, light-blue and green to $\nu_1$ and $\mu_1$, yellow and red to $\nu_2$ and $\mu_2$ in \eqref{eq:pi_splitting}. Each colored region contributes a number of pairings given by the Catalan number indicated. $(b)$ Example of allowed (green) and not-allowed (red) pairs for pairings in $\nc_2^{2}(n,m)$ with mixed pairs given by $(i_1,j_1)$ and $(i_2,j_2)$.}
    \label{fig:help_proof}
\end{figure}

Let us fix $\ell=2k$, and consider $\{(i_a,j_a)\}_{a=1,\ldots,2k}$ a set of mixed pairs $(\circ,\bullet)$. We denote by $\textlmss{NC}_2(\{(i_a,j_a)\}_{a})\subseteq\nc_2^{2k}(2u,2v)$, the set of all pairings $\pi\in\nc_2^{2k}(2u,2v)$ assuming $\{(i_a,j_a)\}_{a}$ as their only mixed-pairs. Furthermore, we will order the pairs such that 
\begin{align}
    1\leq i_1<\cdots<i_{2k}\leq 2u,\quad 1\leq j_1<\cdots<j_{2k}\leq 2v.
\end{align}
Note the strict inequalities since elements cannot have more than one neighbor in the pairing $\pi$. Tautologically, if a pair $(r,s)\in\pi$ is different from all $\{(i_a,j_a)\}_{a}$, it is a pair of type $(\circ,\circ)$ or $(\bullet,\bullet)$.

Take the mixed pair $(i_1,j_1)$. Because of the non-crossing nature of $\pi$, this mixed pair effectively isolates the set $\{j_1-1,\ldots,1,1,\ldots, i_1-1\}$ from all other elements. Therefore, these elements will have to be paired amongst themselves, all the while respecting the non-crossing constraint, i.e. they must create a pairing $\pi_0$ belonging to $\nc_2(i_1-1, j_1-1)$. However, since $\pi_0$ is a sub-pairing of $\pi$, the only mixed pairs allowed for $\pi_0$ are $\{(i_a,j_a)\}_{a=1,\ldots,2k}$, none of which are possible (see \figref{fig:help_proof}-$(b)$ in the case $\ell=2$). Consequently, $\pi_0$ actually factorizes into two non-interacting pairings $\pi_0=\nu_0\sqcup\mu_0$ such that $\nu_0\in\nc_2(j_1-1)$ and $\mu_0\in\nc_2(i_1-1)$. We therefore see that, for $(i_1,j_1)$ to be an admissible mixed-pair, i.e. allowing both $\nc_2(j_1-1)\neq\emptyset$ and $\nc_2(i_1-1)\neq\emptyset$, $i_1$ and $j_1$ must be odd\footnote{Note that in the case where either $i_1=1$ or $j_1=1$, we conventionally write $\nu_0=\emptyset$ or $\mu_0=\emptyset$.}. This reasoning can be repeated inductively to prove that a mixed-pair $(i_a,j_a)$ is possible if $i_a\equiv a\,[2]$ as well as $j_a\equiv a\,[2]$.

To each mixed-pair are associated two sub-pairings $\nu_{a-1},\mu_{a-1}\in\nc_2(i_{a}-i_{a-1}-1),\nc_2(j_{a}-j_{a-1}-1)$, with the convention $i_0=j_0=0$. There are still $n-i_{2k}$ $\circ$-elements and $m-j_{2k}$ $\bullet$-elements to pair (see \figref{fig:help_proof}-$(a)$), which are going to yield two last sub-pairings $\nu_{2k},\mu_{2k}\in\nc_2(i_{2k+1}-i_{2k}-1),\nc_2(j_{2k+1}-j_{2k}-1)$, with the convention $i_{2k+1}=n+1$ and $j_{2k+1}=m+1$. Finally, a pairing $\pi\in\textlmss{NC}_2(\{(i_a,j_a)\}_{a})$ naturally splits into non interacting sub-pairings
\begin{equation}
    \pi=(\nu_0\sqcup\mu_0)\bigsqcup_{a=1}^{2k}(\nu_a\sqcup\mu_a\sqcup(i_a,j_a)).
    \label{eq:pi_splitting}
\end{equation}

The previous reasoning allows us to enumerate the number of pairings $\pi\in\textlmss{NC}_2(\{(i_a,j_a)\}_{a})$. Indeed, using \eqref{eq:pi_splitting}, each $(\nu_a,\mu_a)$ have a total number $N_a$ of possibilities given by
\begin{equation}
    N_a=\card{\nc_2(i_{a+1}-i_a-1)}\card{\nc_2(j_{a+1}-j_a-1)}=C_{\frac{i_{a+1}-i_a-1}{2}}C_{\frac{j_{a+1}-j_a-1}{2}}.
\end{equation}
Therefore, the cardinality of the set $N(\{(i_a,j_a)\}_{a})$ is expressed as 
\begin{equation}
    |N(\{(i_a,j_a)\}_{a})|=\prod_{a=0}^{2k}N_a.
\end{equation}

Now, since no pairing can have two distinct sets of mixed pairs, we have the following natural partitioning of $\nc_2^{2k}(2u,2v)$ into
\begin{equation}
    \nc_2^{2k}(2u,2v)=\bigsqcup_{\textrm{admissible }\{(i_a,j_a)\}_a}\textlmss{NC}_2(\{(i_a,j_a)\}_{a}).
\end{equation}
From the above reasoning, sets admissible mixed-pairs $\{(i_a,j_a)\}_{a}$ are such that $i_a,j_a$ both have the same parity as $a$. We can therefore express the cardinality of $\nc_2^{2k}(2u,2v)$ as
\begin{align*}
    \card{\nc_2^{2k}(2u,2v)}&=\sum_{\textrm{admissible }\{(i_a,j_a)\}_a}\card{N(\{(i_a,j_a)\}_{a})}\\
    &= \sum_{\substack{1\leq i_1<\cdots<i_{2k}\leq 2u\\1\leq j_1<\cdots<j_{2k}\leq 2v\\ i_a\equiv a\,[2],\,j_a\equiv a\,[2]}}\prod_{a=0}^{2k} N_a\\
    &= \left(\sum_{\substack{1\leq i_1<\cdots<i_{2k}\leq 2u\\ i_a\equiv a\,[2]}}\prod_{a=0}^{2k} C_{\frac{i_{a+1}-i_a-1}{2}}\right)\left(\sum_{\substack{1\leq j_1<\cdots<j_{2k}\leq 2v\\ j_a\equiv a\,[2]}}\prod_{a=0}^{2k} C_{\frac{j_{a+1}-j_a-1}{2}}\right)
\end{align*}
which we rewrite as
\begin{equation}
    \card{NC_2^{2k}(2u,2v)}:=B(k,u)B(k,v),
    \label{eq:prod_cardinal}
    \end{equation}
where $B(x,y)$ is 
\begin{equation}
    B(x,y)=\sum_{\substack{1\leq i_1<\cdots<i_{2x}\leq 2y\\ i_a\equiv a\,[2]}}\prod_{a=0}^{2x} C_{\frac{i_{a+1}-i_a-1}{2}}.
    \label{eq:def_B}
\end{equation}

Let us now express $B(k,t)$ for any $t$. By de-nesting the sums over indices $i_{2k}$ and $i_{2k-1}$, we have
\begin{align}
    B(k,t)&=\sum_{\substack{i_{2k}=2k\\i_{2k}\equiv 0\,[2]}}^{2t}C_{\frac{2t-i_{2k}}{2}}\sum_{\substack{i_{2k-1}=2k-1\\i_{2k-1}\equiv 1\,[2]}}^{i_{2k}-1}C_{\frac{i_{2k}-i_{2k-1}-1}{2}}\left(\sum_{\substack{1\leq i_1<\cdots<i_{2k-2}\leq i_{2k-1}-1\\ i_a\equiv a\,[2]}}\prod_{a=0}^{2k-2} C_{\frac{i_{a+1}-i_a-1}{2}}\right)\nonumber\\
    &=\sum_{\substack{i_{2k}=2k\\i_{2k}\equiv 0\,[2]}}^{2t}C_{\frac{2t-i_{2k}}{2}}\sum_{\substack{i_{2k-1}=2k-1\\i_{2k-1}\equiv 1\,[2]}}^{i_{2k}-1}C_{\frac{i_{2k}-i_{2k-1}-1}{2}}B\left(k-1,\frac{i_{2k-1}-1}{2}\right).\nonumber
    \intertext{Upon the change of summation indices $i_{2k}=2r$, $i_{2k-1}=2s-1$, which enforces the parity constraints, we get a recursive formula for $B(\cdot,\cdot)$}
    B(k,t)&=\sum_{r=k}^{t}C_{t-r}\sum_{s=k}^{r}C_{r-s}B(k-1,s-1).\label{eq:rec_B}
\end{align}
From this recursive formula, one can show (see Appendix \ref{ap:catalan_triangle} and \cite{AIGNER199933, carlitz1972sequences}) that $B(k,t)$ can be expressed as a particular Catalan  triangular number $C(x,y)$ (or ballot number, see \cite{AIGNER199933}), i.e.
\begin{equation}
    B(k,t)=C(t+k+1,t-k+1)=\frac{2k+1}{t+k+1}\binom{2t}{t+k}.
\end{equation}
Therefore, using \eqref{eq:prod_cardinal} the cardinality of $\nc_2^{2k}(2u,2v)$ can be written as \eqref{eq:objective_combinatorics}:
\begin{equation}
    \card{\nc_2^{2k}(2u,2v)}=\frac{(2k+1)^2}{(u+k+1)(v+k+1)}\binom{2u}{u+k}\binom{2v}{v+k}.
\end{equation}

\subsection{\texorpdfstring{Cardinality of $\nc^{2k}_2(t,s)$: mapping to Temperley-Lieb diagrams}{}}

There is however a quicker way to study $\nc^{2k}_2(t,s)$. Consider a pairing $\pi\in\nc_2(2u,2v)$ which we represent on a line, i.e. for $u=2$ and $v=1$, we can have for instance:
\begin{equation*}
\begin{tikzpicture}[scale=0.6, every node/.style={scale=1}]
    \draw (5,-5) to[bend left=60] (10,-5);
    \draw (6,-5) to[bend left=60] (7,-5);
    \draw (8,-5) to[bend left=60] (9,-5);
    
    \draw (5,-5)  node[ptm={white}] {};
    \draw (6,-5) node[ptm={white}] {};
    \draw (7,-5) node[ptm={white}] {};
    \draw (8,-5) node[ptm={white}] {};
    \draw (9,-5) node[ptm={black}] {};
    \draw (10,-5) node[ptm={black}] {};
%     \node[font=\Large] at (11.5,-5) {$\xrightleftharpoons[\varphi^{-1}]{\varphi}$};

% \begin{scope}[xshift=0.5cm]
%      \draw (12,-6) [dnup={1}{white}{black}];
%     \draw (13,-6) [dndn={1}{white}{white}];
%     \draw (15,-6) [dnup={-1}{white}{black}];
% \end{scope}
\end{tikzpicture}.
%\label{eq:mapping_nc_tl}
\end{equation*}
Now, consider the following move: take the right-most $\bullet$ element and move it up so that it hovers over the others, i.e.
\begin{equation*}
\begin{tikzpicture}[scale=0.6, every node/.style={scale=1}]
    \draw (6,-5) to[bend left=60] (7,-5);
    \draw (8,-5) to[bend left=60] (9,-5);
    
    \draw (6,-5) node[ptm={white}] {};
    \draw (7,-5) node[ptm={white}] {};
    \draw (8,-5) node[ptm={white}] {};
    \draw (9,-5) node[ptm={black}] {};
    \draw (5,-5) [dnup={1}{white}{black}];
\end{tikzpicture}
\end{equation*}
Repeat the same move with the new right-most $\bullet$ element at the bottom:
\begin{equation*}
\begin{tikzpicture}[scale=0.6, every node/.style={scale=1}]
    \draw (6,-5) to[bend left=60] (7,-5);

    \draw (6,-5) node[ptm={white}] {};
    \draw (7,-5) node[ptm={white}] {};

    \draw (5,-5) [dnup={1}{white}{black}];
    \draw (8,-5) [dnup={-1}{white}{black}];
\end{tikzpicture}
\end{equation*}
For an arbitrary paring $\pi$, this sequence of $2v$ moves defines a mapping $\varphi_{2v}$ from two topologically equivalent structures embedded in the plane
\begin{equation}
\begin{aligned}
\begin{tikzpicture}[scale=0.6, every node/.style={scale=1}]
    \draw (5,-5) to[bend left=60] (10,-5);
    \draw (6,-5) to[bend left=60] (7,-5);
    \draw (8,-5) to[bend left=60] (9,-5);
    
    \draw (5,-5)  node[ptm={white}] {};
    \draw (6,-5) node[ptm={white}] {};
    \draw (7,-5) node[ptm={white}] {};
    \draw (8,-5) node[ptm={white}] {};
    \draw (9,-5) node[ptm={black}] {};
    \draw (10,-5) node[ptm={black}] {};
    \node[font=\Large] at (11.5,-5) {$\xrightleftharpoons[\varphi^{-1}_{2}]{\varphi_2}$};
\begin{scope}[xshift=8cm]
     \draw (6,-5) to[bend left=60] (7,-5);

    \draw (6,-5) node[ptm={white}] {};
    \draw (7,-5) node[ptm={white}] {};

    \draw (5,-5) [dnup={1}{white}{black}];
    \draw (8,-5) [dnup={-1}{white}{black}];
\end{scope}
\end{tikzpicture}
\end{aligned}.
\label{eq:mapping_nc_tl}
\end{equation}
The resulting structure $\varphi_{2v}(\pi)$ is called a Temperley-Lieb diagram, belonging to the Temperley-Lieb algebra $\mathcal{TL}_{2u,2v}$. These diagrams connect $2u$ $\circ$-elements with $2v$ $\bullet$-elements spread over two opposite strands (connections between elements of the same strand are allowed). Moreover, the connection pattern has a non-crossing structure. The \emph{rank} of the diagram corresponds to the number of connections between elements belonging to different sides. It is straightforward to see that, the isomorphic mapping, geometrically described in \eqref{eq:mapping_nc_tl}, sends elements from $\nc^{2k}_2(2u,2v)$, with $\sigma_c=2k$, onto diagrams with rank $2k$. Finally, it is known that the rank function furnishes a simple way to partition $\mathcal{TL}_{2u,2v}$ into constant-rank subsets $\mathcal{TL}^{2k}_{2u,2v}$ and that these subsets have a cardinality given by products of triangular Catalan numbers (see \cite{TLAlgebras}). The cardinality of $\nc^{2k}_2(2u,2v)$ is therefore given by
\begin{equation}
        \card{\nc^{2k}_2(2u,2v)}=\card{\mathcal{TL}^{2k}_2(2u,2v)}=\frac{(2k+1)^2}{(u+k+1)(v+k+1)}\binom{2u}{u+k}\binom{2v}{v+k}.
\end{equation}

\subsection{\texorpdfstring{Polynomial expression for $\mm{n}{m}(\rho)$}{}}

We can now state the main result of this paper by applying the results of the previous analysis to \eqref{eq:poly_formula}. If $n=2u$ and $m=2v$, the mixed moment $\mm{2u}{2v}(\rho)$ has the explicit formula
\begin{subequations}
\begin{align}
\mm{2u}{2v}(\rho)&=\sum_{k=0}^{u\wedge v}\rho^{u+v-2k}\frac{(2k+1)^2}{(u+k+1)(v+k+1)}\binom{2u}{u+k}\binom{2v}{v+k},\label{eq:explicit_formulas_even}\\
\intertext{whereas when $n=2u+1$ and $m=2v+1$ are odd, we get}
\mm{2u+1}{2v+1}(\rho)&=\sum_{k=0}^{u\wedge v}\rho^{u+v-2k}\frac{4(k+1)^2}{(u+k+2)(v+k+2)}\binom{2u+1}{u+k+1}\binom{2v+1}{v+k+1}.\label{eq:explicit_formulas_odd}
\end{align}
\label{eq:explicit_formulas}
\end{subequations}
Again, if $n,m$ are not of the same parity, $\mm{n}{m}(\rho)=0$ since there are no pairings over an odd number of elements. One can see that \eqref{eq:explicit_formulas_even} and \eqref{eq:explicit_formulas_odd} are symmetric upon the exchange $u\leftrightarrow v$, which accounts for the invariance of $\tr{\cdot}$ under both cyclic permutation and transposition. Let us illustrate this formula with some examples. For $n=6$ and $m=2$, our formula gives $\mm{6}{2}(\rho)=5\rho^4+9\rho^2$, which we can check with an explicit enumeration
\begin{equation*}
    \begin{aligned}
\tikz[scale=0.6, every node/.style={scale=1}]{\node at (0,0) {};
\node at (0.0,0.95) {$\mm{6}{2}(\rho)$}}&
\tikz[blue, scale=0.6, every node/.style={scale=1, black}]{
\node at (-1.65,0) {=};
\CircleMatching{8}{1/8/white/black,2/7/white/black,3/6/white/white,4/5/white/white}{1}{0,0};}
\;
\tikz[blue, scale=0.6, every node/.style={scale=1, black}]{
\node at (-1.65,0) {+};
\CircleMatching{8}{1/8/white/black,2/7/white/black,3/4/white/white,5/6/white/white}{1}{0,0};}\;
\tikz[blue, scale=0.6, every node/.style={scale=1, black}]{
\node at (-1.65,0) {+};
\CircleMatching{8}{1/8/white/black,2/3/white/white,4/7/white/black,5/6/white/white}{1}{0,0};}
\;
\tikz[blue, scale=0.6, every node/.style={scale=1, black}]{
\node at (-1.65,0) {+};
\CircleMatching{8}{1/8/white/black,2/3/white/white,4/5/white/white,6/7/white/black}{1}{0,0};}
\;
\tikz[blue, scale=0.6, every node/.style={scale=1, black}]{
\node at (-1.65,0) {+};
\CircleMatching{8}{1/8/white/black,2/5/white/white,3/4/white/white,6/7/white/black}{1}{0,0};}
\;
\tikz[blue, scale=0.6, every node/.style={scale=1, black}]{
\node at (-1.65,0) {+};
\CircleMatching{8}{1/2/white/white,3/8/white/black,4/7/white/black,5/6/white/white}{1}{0,0};}
\;
\tikz[blue, scale=0.6, every node/.style={scale=1, black}]{
\node at (-1.65,0) {+};
\CircleMatching{8}{1/2/white/white,3/8/white/black,4/5/white/white,6/7/white/black}{1}{0,0};}
\;
\tikz[scale=0.6, every node/.style={scale=1}]{\node at (0,0) {};
\node[label={[xshift=0.cm, yshift=-0.4cm]{$5\rho^4+9\rho^2$,}}] at (0,0.95) {};
\node at (-1.65,0.95) {=};}\\
&
\tikz[blue, scale=0.6, every node/.style={scale=1, black}]{
\node at (-1.65,0) {\phantom{=}};
\CircleMatching{8}{1/2/white/white,3/4/white/white,5/8/white/black,6/7/white/black}{1}{0,0};}
\;
\tikz[blue, scale=0.6, every node/.style={scale=1, black}]{
\node at (-1.65,0) {+};
\CircleMatching{8}{1/4/white/white,2/3/white/white,5/8/white/black,6/7/white/black}{1}{0,0};}
\;
\tikz[teal, scale=0.6, every node/.style={scale=1, black}]{
\node at (-1.65,0) {+};
\CircleMatching{8}{1/6/white/white,2/5/white/white,3/4/white/white,7/8/black/black}{1}{0,0};}
\;
\tikz[teal, scale=0.6, every node/.style={scale=1, black}]{
\node at (-1.65,0) {+};
\CircleMatching{8}{1/6/white/white,2/3/white/white,4/5/white/white,7/8/black/black}{1}{0,0};}
\;
\tikz[teal, scale=0.6, every node/.style={scale=1, black}]{
\node at (-1.65,0) {+};
\CircleMatching{8}{1/2/white/white,3/6/white/white,4/5/white/white,7/8/black/black}{1}{0,0};}
\;
\tikz[teal, scale=0.6, every node/.style={scale=1, black}]{
\node at (-1.65,0) {+};
\CircleMatching{8}{1/4/white/white,2/3/white/white,5/6/white/white,7/8/black/black}{1}{0,0};}
\;
\tikz[teal, scale=0.6, every node/.style={scale=1, black}]{
\node at (-1.65,0) {+};
\CircleMatching{8}{1/2/white/white,3/4/white/white,5/6/white/white,7/8/black/black}{1}{0,0};}
\end{aligned}
\label{eq:example_p_6_2}
\end{equation*}
where we highlighted in blue the pairings contributing to $\rho^2$, i.e. with two mixed pairs, and in green those contributing to $\rho^4$, i.e. with four mixed pairs. We can also check in the odd case $n=5$ and $m=3$, $\mm{5}{3}(\rho)=10\rho^3+4\rho$, which we confirm by enumeration
 
\begin{equation*}
    \begin{aligned}
\tikz[scale=0.6, every node/.style={scale=1}]{\node at (0,0) {};
\node at (0.0,0.95) {$\mm{5}{3}(\rho)$}}&
\tikz[cyan, scale=0.6, every node/.style={scale=1, black}]{
\node at (-1.65,0) {=};
\CircleMatching{8}{1/8/white/black,2/7/white/black,3/6/white/black,4/5/white/white}{1}{0,0};}
\;
\tikz[cyan, scale=0.6, every node/.style={scale=1, black}]{
\node at (-1.65,0) {+};
\CircleMatching{8}{1/8/white/black,2/7/white/black,3/4/white/white,5/6/white/black}{1}{0,0};}\;
\tikz[cyan, scale=0.6, every node/.style={scale=1, black}]{
\node at (-1.65,0) {+};
\CircleMatching{8}{1/8/white/black,2/3/white/white,4/7/white/black,5/6/white/black}{1}{0,0};}
\;
\tikz[red, scale=0.6, every node/.style={scale=1, black}]{
\node at (-1.65,0) {+};
\CircleMatching{8}{1/8/white/black,2/3/white/white,4/5/white/white,6/7/black/black}{1}{0,0};}
\;
\tikz[red, scale=0.6, every node/.style={scale=1, black}]{
\node at (-1.65,0) {+};
\CircleMatching{8}{1/8/white/black,2/5/white/white,3/4/white/white,6/7/black/black}{1}{0,0};}
\;
\tikz[cyan, scale=0.6, every node/.style={scale=1, black}]{
\node at (-1.65,0) {+};
\CircleMatching{8}{1/2/white/white,3/8/white/black,4/7/white/black,5/6/white/black}{1}{0,0};}
\;
\tikz[red, scale=0.6, every node/.style={scale=1, black}]{
\node at (-1.65,0) {+};
\CircleMatching{8}{1/2/white/white,3/8/white/black,4/5/white/white,6/7/black/black}{1}{0,0};}
\;
\tikz[scale=0.6, every node/.style={scale=1}]{\node at (0,0) {};
\node[label={[xshift=0.cm, yshift=-0.4cm]{$10\rho^3+4\rho$,}}] at (0,0.95) {};
\node at (-1.65,0.95) {=};}\\
&
\tikz[red, scale=0.6, every node/.style={scale=1, black}]{
\node at (-1.65,0) {\phantom{=}};
\CircleMatching{8}{1/2/white/white,3/4/white/white,5/8/white/black,6/7/black/black}{1}{0,0};}
\;
\tikz[red, scale=0.6, every node/.style={scale=1, black}]{
\node at (-1.65,0) {+};
\CircleMatching{8}{1/4/white/white,2/3/white/white,5/8/white/black,6/7/black/black}{1}{0,0};}
\;
\tikz[red, scale=0.6, every node/.style={scale=1, black}]{
\node at (-1.65,0) {+};
\CircleMatching{8}{1/6/white/black,2/5/white/white,3/4/white/white,7/8/black/black}{1}{0,0};}
\;
\tikz[red, scale=0.6, every node/.style={scale=1, black}]{
\node at (-1.65,0) {+};
\CircleMatching{8}{1/6/white/black,2/3/white/white,4/5/white/white,7/8/black/black}{1}{0,0};}
\;
\tikz[red, scale=0.6, every node/.style={scale=1, black}]{
\node at (-1.65,0) {+};
\CircleMatching{8}{1/2/white/white,3/6/white/black,4/5/white/white,7/8/black/black}{1}{0,0};}
\;
\tikz[red, scale=0.6, every node/.style={scale=1, black}]{
\node at (-1.65,0) {+};
\CircleMatching{8}{1/4/white/white,2/3/white/white,5/6/white/black,7/8/black/black}{1}{0,0};}
\;
\tikz[red, scale=0.6, every node/.style={scale=1, black}]{
\node at (-1.65,0) {+};
\CircleMatching{8}{1/2/white/white,3/4/white/white,5/6/white/black,7/8/black/black}{1}{0,0};}
\end{aligned}
\label{eq:example_p_5_3}
\end{equation*}
and where we highlighted in red the pairings contributing to $\rho$, i.e. with one mixed pair, and in green those contributing to $\rho^3$, i.e. with three mixed pairs. Of course, it is still possible to enumerate all pairings in the above cases and check our formula. In Appendix \ref{ap:p_4_8}, we show, on a specific example for which $n=4$ and $m=8$, that the enumeration quickly becomes unmanageable. We can readily see the gain in computing power allowed by our formula.

\begin{figure}[!tb]
    \centering
    \includegraphics{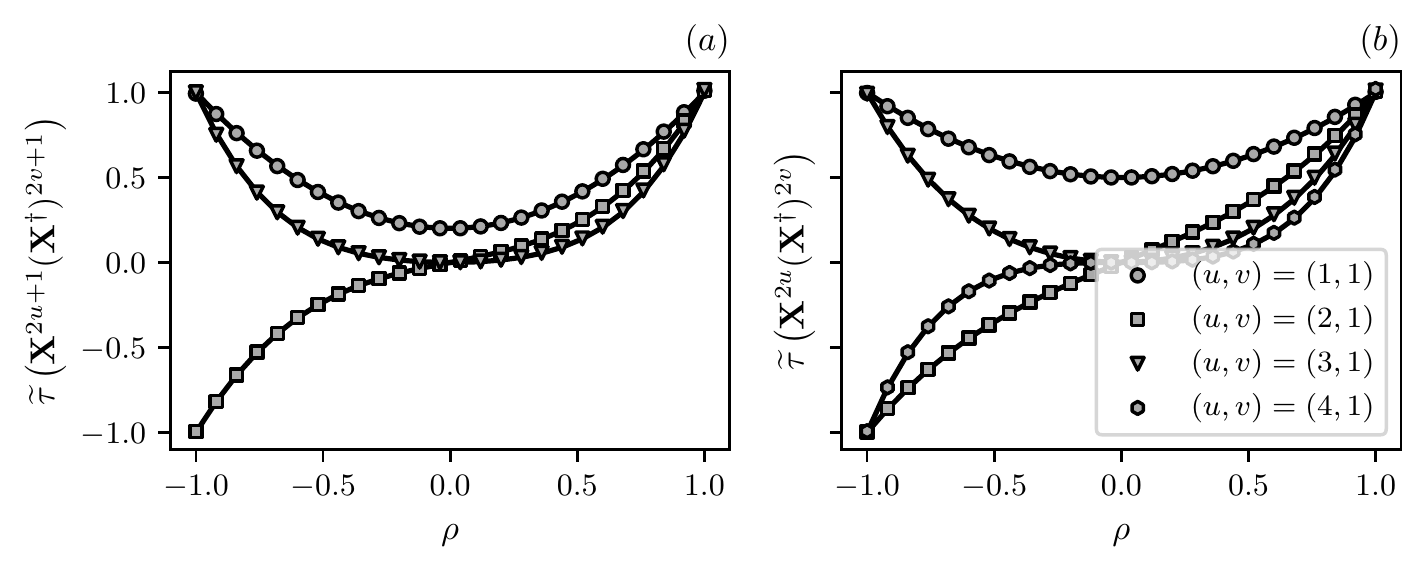}
    \caption{Comparison between $\widetilde{\tau}\left(\X^n(\Xt)^m\right)=\tau\left(\X^n(\Xt)^m\right)/C_{(n+m)/2}$ (dark gray points) and $\widetilde{\mathcal{O}}_{n}^{m}=\mm{n}{m}/C_{(n+m)/2}$ (solid black lines) for $\rho\in[-1,1]$. $(a)$ Odd $n=2u+1,m=2v+1$, $(b)$ Even $n=2u,m=2v$. For each point, we simulated $k=100$ elliptic matrices of size $500\times500$ and computed the average of the normalized trace of the product $\X^n(\X^\dagger)^m$. We see an excellent agreement between simulations and theoretical values.}
    \label{fig:th_vs_exp}
\end{figure}

Finally, even though \eqref{eq:explicit_formulas_even} and \eqref{eq:explicit_formulas_odd} coincide with \eqref{eq:final_formula}, one must check that it coincides with the trace $\tr{\X^n(\Xt)^m}$ averaged over matrices $\X$ drawn from \gee. On \figref{fig:th_vs_exp} we show the excellent agreement between our theoretical formula and numerical simulations

\subsection{\texorpdfstring{Special values for $\rho$}{}}

One must check that $\mm{n}{m}(1)=C_{(n+m)/2}$ to retrieve the well-known result concerning the moments of \goe~matrices, see for instance \cite{potters_bouchaud_2020}. Using results from \cite{AIGNER199933} (or using the mapping onto Temperley-Lieb diagrams), we see that both identities
\begin{equation}
\begin{aligned}
    \sum_{k=0}^{u\wedge v}\frac{(2k+1)^2}{(u+k+1)(v+k+1)}\binom{2u}{u+k}\binom{2v}{v+k}&=C_{u+v}\\ 
    \sum_{k=0}^{u\wedge v}\frac{4(k+1)^2}{(u+k+2)(v+k+2)}\binom{2u+1}{u+k+1}\binom{2v+1}{v+k+1}&=C_{u+v+1} 
\end{aligned}
\end{equation}
hold, and we get the expected result $\mm{n}{m}(1)=C_{(n+m)/2}$. Furthermore, it is easy to see that the parity of the polynomial function $\mm{n}{m}$ depends on the parity of $n$, $m$, and $(n+m)/2$. Indeed, on the one hand, assuming $n,m\in2\N$, powers of $\rho$ will have the same parity as $(n+m)/2$. On the other hand if $n,m\in2\N+1$, they will have an opposite parity from $(m+n)/2$. As a consequence, we retrieve the moments of anti-\goe~matrices
\begin{equation}
P_{n,m}(-1)=\begin{dcases}
(-1)^{u+v}C_{u+v}\,, &\text{for $n=2u,\,m=2v$}\\
(-1)^{u+v}C_{u+v+1}\,, &\text{for $n=2u+1,\,m=2v+1$}
\end{dcases}.
\end{equation}
One can quickly check that these results are coherent with a direct computation of $\tr{\X^{n}(\Xt)^{m}}$ for $\rho=\pm1$.

As mentioned above, considering that $\X$ is a Ginibre matrix, i.e. $\rho=0$, the computation of $\mm{n}{m}(0)$ should give the answer to the Knights and Ladies problem (see \cite{kemp_enumeration_2011}). In our simplified setting, it is easy to see that plugging $\rho=0$ in $\mm{n}{m}$ always yields $0$ (there are no solutions to the problem), except when $n=m$ where the only possible pairing is the one systematically connecting $\circ$-elements to $\bullet$-elements, e.g.
\begin{equation*}
\begin{aligned}
\begin{tikzpicture}[scale=0.6, every node/.style={scale=1}]
\CircleMatching{12}{1/12/white/black,
                    2/11/white/black,
                    3/10/white/black,
                    4/9/white/black,
                    5/8/white/black,
                    6/7/white/black}{1}{0,0};
\end{tikzpicture}
\end{aligned}
\end{equation*}
in the case $n=m=6$. As a consequence
\begin{equation}
\mm{n}{m}(0)=\delta_{nm},
\end{equation}
with $\delta_{ij}=1$ whenever $i=j$ and $0$ otherwise. This result is reported in \cite{kemp_enumeration_2011}. In the context of the cone-wise linear dynamics from \cite{Dessertaine2022Cones}, this result shows that, in the Ginibre case, the Gaussian process on which the dynamical evolution can be mapped is completely uncorrelated both in space and time. The distribution $q(\tau)$ of the time spent in either region, therefore, decays exponentially $q(\tau)=2^{-\tau}$.

\section{\texorpdfstring{Asymptotics of $
\mm{n}{m}(\rho)$ as $n,m\to\infty$}{}\label{sec:sec3}}

Without loss of generality, we can consider the case where $\rho\geq0$\footnote{In \eqref{eq:explicit_formulas}, we can rewrite $\rho^{u+v-2k}$ as $(-1)^{u+v}|\rho|^{u+v-2k}$ and study the asymptotic behavior of the sums replacing $\rho$ by $|\rho|$.} and study the asymptotic behavior of $\mm{n}{m}$ in the limit of a large number of terms in the mixed-moment. Furthermore, we will consider the case where both $n=2u$ and $m=2v$ are even such that $m\leq n$. In the context of \cite{Dessertaine2022Cones}, the asymptotic behavior of $\mm{n}{m}(\rho)$ determines the asymptotic behavior of the covariance, which, in some cases, can be an indicator of the asymptotic behavior of the persistence (see \cite{BrayReviewPersistence,NewellRosenblatt1962}). As computations are quite lengthy, we will only give the results here, but see Appendix \ref{ap:asymptotic} for the details. Furthermore, we will consider rescaled mixed-moments $\rmm{n}{m}$ defined as $\rmm{n}{m}:=\mm{n}{m}/\sqrt{\mm{n}{n}\mm{m}{m}}$, and such that $\rmm{n}{m}\leq1$.
\begin{figure}[b]
    \centering
    \includegraphics{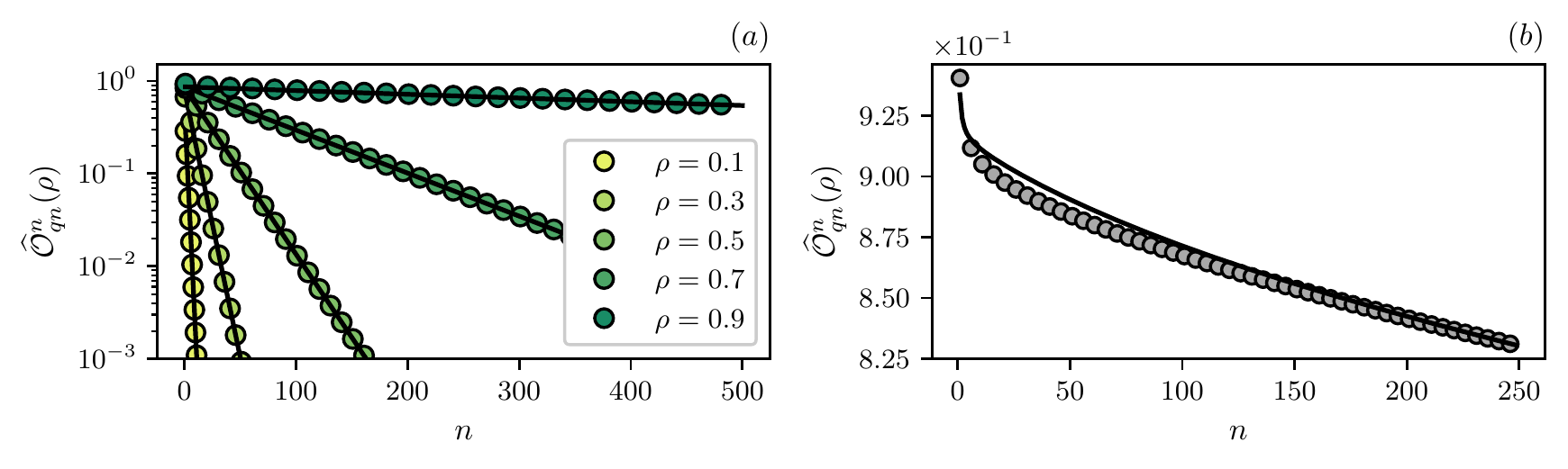}
    \caption{$(a)$ Rescaled polynomial $\widehat{\mathcal{O}}_{qn}^n(\rho):=\mm{qn}{n}(\rho)/\sqrt{\mm{qn}{qn}(\rho)\mm{n}{n}(\rho)}$ (colored points) as a function of $n$, different values of $0<\rho<1$ and $q=2$. We indicated the asymptotic estimate in the solid black lines. $(b)$ Zoom onto the region $n\leq 250$ for $\rho=0.95$. We see that the error between the two plots decreases as $n$ grows larger.}
    \label{fig:correlator_elliptic_decay}
\end{figure}

Whenever $\rho=1$, we know that $\mm{2u}{2v}(1)=C_{u+v}$, which, as $u,v\to\infty$ gives the asymptotic behavior 
\begin{equation}
    \rmm{2u}{2v}(1)\underset{u,v\to\infty}{\sim}\left(\frac{2\sqrt{uv}}{u+v}\right)^{3/2}.
\end{equation}
If we consider $u$ and $v$ such that $u=qv$ for some fixed $q\geq1$, the previous expression assumes a finite limit
\begin{equation}
    \rmm{2qv}{2v}(1)\underset{u,v\to\infty}{\sim}\left(\frac{2\sqrt{q}}{1+q}\right)^{3/2}.
\end{equation}
This scaling turns out to be the correct one to study the asymptotics of $\rmm{n}{m}(\rho)$ for any $\rho$. Using results from Appendix \ref{ap:asymptotic}, we get
\begin{equation}
    \rmm{2qv}{2v}(\rho)\underset{u,v\to\infty}{\sim}\Psi_q(\rho)e^{-v\Phi_q(\rho)},
\end{equation}
where $\Psi_q$ and $\Phi_q$ are given in Appendix \ref{ap:asymptotic}. At fixed $\rho$, note that both $\Psi_q$ and $\Phi_q$ only depend on the specific ray of parameters $(qv,v)$ that we consider. Using the fact that $v=(u+v)/(q+1)$, one could write the asymptotics for any $u,v$ as
\begin{equation}
    \rmm{2u}{2v}(\rho)\underset{u,v\to\infty}{\sim}\Psi_{\frac{u}{v}}(\rho)e^{-(u+v)\hat{\Phi}_{\frac{u}{v}}(\rho)},
    \label{eq:final_asymptotics}
\end{equation}
where $\Psi$ and $\hat{\Phi}=\Phi/(q+1)$ are self-similar functions in $u,v$. On \figref{fig:correlator_elliptic_decay}-$(a)$, we show the agreement between $\rmm{qn}{n}$ and the estimate in \eqref{eq:final_asymptotics}. Furthermore, in Appendix \ref{ap:asymptotic}, we show how to recover the asymptotic behavior of the Catalan numbers in the limit $\rho\to1^-$.

\section{\texorpdfstring{Some more general mixed-moments}{}\label{sec:sec4}}

In this final section, we will consider more general mixed-moments of \gee~matrices. Instead of grouping matrices in blocks of $\X$ or $\Xt$ as in the previous section, we will record the positions of matrices $\X$ in the normalized trace. We will denote by $\i$ the tuple encoding these positions in the product (read from left to right) of $2M$ terms $\X^{\eps_1}\cdots\X^{\eps_{2M}}$ with $\eps_i\in\{1,\dagger\}$. Furthermore, we will consider $\i$ to be ordered such that
\begin{equation}
    1\leq i_1<i_2<\cdots<i_k\leq 2M,
\end{equation}
whenever there are $k$ matrices $\X$ in the previous product. We denote by $\P(\i)$ the product associated with $\i$. For instance, with $M=4$, $\i=(1,2,3,7)\rightarrow\X^3(\Xt)^3\X\Xt=\P(\i)$. In the following, we will only consider tuples $\i$ where there are more odd elements than even ones. This does not limit the scope of the analysis since $\varphi(\cdot)$ is invariant under cyclic permutation~\footnote{Indeed, considering $R_1$ the transformation of $\i$ sending every indices $i_\ell$ to $i_{\ell}+1$ with periodic boundary condition (i.e $i_{k}=2M$ is sent to $1$), we have 
\[\varphi(\P(\i))=\varphi(\P(R_1(\i))).\]
Now, since the number of odd elements in $\i$ is equal to the number of even elements in $R_1(\i)$ we can restrict ourselves to the aforementioned case.}. Furthermore, we will also restrict ourselves to tuples $\i$ of length $k\leq M$ since $\varphi(\cdot)$ is invariant under the exchange $\X\leftrightarrow\Xt$~\footnote{In \eqref{eq:final_formula}, $\sigma$ is invariant under the exchange $\X\leftrightarrow\Xt$ since mixed-pairs $(\circ,\bullet)$ will remain mixed under this transformation. Now, if the length of $\i$ is such that $k>M$, one only has to exchange $\X$ and $\Xt$ to retrieve the case $k\leq M$.}.

\subsection{General case}

For a particular pairing $\pi$, it is not hard to see that the quantity $\sigma(\pi)$ from \eqref{eq:final_formula} is completely determined by the number of pairs $(\circ,\circ)$ (pairs linking two $\X$'s) that are formed. Denoting by $\sigma_\circ$ the function $\sigma_{\circ}:\pi\mapsto\card{\{(r,s)\in\pi,(\circ,\circ)=(r,s)\}}$, we have the relationship $\sigma(\pi)=2\sigma_\circ(\pi)+M-k$, where $k$ is the length of $\i$. Therefore, it is enough to study $\sigma_\circ$ to express \eqref{eq:final_formula}. Finally, any pair in $\pi$ must be formed by an even element and an odd element to preserve the non-crossing structure of the pairing (see \secref{sec:sec3} for the details of this argument).

For an even element $2e\in\i$ and an odd element $2\ell-1\in\i$, we will denote by $\Ac_{e,\ell}$ the sets $\Ac_{e,\ell}=\{\pi\in\nc_2(2M),\,(2e,2\ell-1)\in\pi\}$ and by $\Att_{e}(\i)$ the set
\begin{equation}
\Att_{e}(\i)=\bigsqcup_{2\ell-1\in\i}\Ac_{e,\ell}.
\end{equation}
It is easy to see that generically $\nc_2(2M)=\bigsqcup_{\ell=1}^M\Ac_{e,\ell}$, since any $\pi\in\nc_2(2M)$ will have an edge either beginning or ending in $2e$, whose other element is bound to be odd. One can therefore express the cardinality of the sets $\Att_e(\i)$ as
\begin{equation}
\card{\Att_e(\i)}=C_M-\sum_{2\ell-1\notin\i}\card{\Ac_{e,\ell}}.
\end{equation}
With the same reasoning as in \secref{sec:sec2}, one can show that
\begin{equation}
\Ac_{e,\ell}\cong
\begin{cases}
\nc_2(2(e-\ell))\times\nc_2(2(M-e+\ell-1))&,\quad \ell\leq e\\
\nc_2(2(\ell-e-1))\times\nc_2(2(M-\ell+e))&,\quad \ell> e.
\end{cases}
\label{eq:bijection_Ac}
\end{equation}
Furthermore we denote by $N_{j}(\i)$ (resp. $N_{\geq j}(\i)$) the cardinality of the sets $\{\pi\in\nc_2(2M),\, \sigma_\circ(\pi)=j\}$ (resp. $\{\pi\in\nc_2(2M),\, \sigma_\circ(\pi)\geq j\}$), where we can link both quantities through $N_j(\i)=N_{\geq j}(\i)-N_{\geq j+1}(\i)$. With the previous reasoning on $\sigma_\circ$, we can rewrite \eqref{eq:final_formula} as 
\begin{equation}
\varphi(\P(\i))=\rho^{M-k}\sum_{j=1}^{r}\rho^{2j}N_{j}(\i),
\label{eq:general_gem}
\end{equation}
where $r$ is the number of even numbers in $\i$. The numbers $N_{\geq j}(\i)$ count the number of pairings with at least $j$ pairs $(\circ,\circ)$ such that
\begin{equation}
N_{\geq j}(\i)=\card{\bigcup_{1\leq k_1<\cdots<k_{j}\leq r}\,\bigcap_{s=1}^{j}\Att_{e_{k_s}}(\i)}.
\label{eq:int_N_geq_j}
\end{equation}
Denoting by $\k$ the ordered tuples $(k_1,\ldots,k_j)\in[\![ 1, r]\!]^j$, and by $\Att_{\k}(\i)$ the sets 
\[\Att_{\k}(\i)=\bigcap_{q\in\k}\Att_{e_{q}}(\i),\]
we can use the inclusion-exclusion principle to further express \eqref{eq:int_N_geq_j} into
\begin{equation}
N_{\geq j}(\i)=\sum_{s=1}^{\binom{r}{j}}(-1)^{s-1}\sum_{\k_1<\ldots<\k_s}\card{\bigcap_{u=1}^s{\Att_{\k_u}(\i)}},
\label{eq:N_geq_j}
\end{equation}
where the order between $j$-tuples refers to the lexicographical order. In full generality, \eqref{eq:N_geq_j} is complicated to track down, and we will therefore focus on the cases $r=0,1,2$ in the following.

\subsection{\texorpdfstring{$r=0$: $\i$ only has odd elements}{}}

In this first very simple case, pairing matrices $\X$ together is impossible since, as we explained in the previous section, one can only pair elements indexed by numbers of opposite parity. Therefore, for any $\pi\in\nc_2(2M)$, $\sigma_\circ(\pi)=0$, which in turn implies $\sigma(\pi)=M-k$. Introducing $o_\ell$ such that $\i=(2 o_1-1,\ldots,2o_k-1)$, we get the mixed-moment
\begin{equation}
    \tr{\P(\i)}=C_M\rho^{M-k},
    \label{eq:final_r_0}
\end{equation}
where we have
\begin{equation}
    \P(\i)=(\Xt)^{2(o_1-1)}(\X\Xt)^{o_2-o_1+1}(\X^\dagger)^{2(o_2-o_3)}\cdots(\X\Xt)^{o_{k}-o_{k-1}+1}(\Xt)^{2(M-o_k)}.
\end{equation}
For example, if $M=8$ and $\i=(1,3,7)$, we have 
\[\tr{(\X\Xt)^2(\Xt)^2\X\Xt}=14\rho,\]
which we can easily check via an explicit enumeration. Note that whenever $k=M$, i.e $\P(\i)=(\X\Xt)^M$, we have $\tr{(\X\Xt)^M}=C_M=FC_1(M)$ as in the Ginibre case.

\subsection{\texorpdfstring{$r=1$: $\i$ has exactly one even element}{}}

Denoting by $2e$ the only even element in $\i$, the maximum of $\sigma_\circ(\pi)$ is now $1$: the only way to pair two matrices $\X$ together is to link $2e$ with any of the $k-1$ odd elements in $\i$. Therefore, $N_1(\i)=N_{\geq 1}(\i)$ and we have 
\begin{equation}
N_1(\i)=\card{\Att_{e}(\i)}=C_M-\sum_{2\ell-1\notin\i}\card{\Ac_{e,\ell}}.
\end{equation}
Using \eqref{eq:bijection_Ac} and the fact that, in this case $N_0(\i)=C_M-N_1(\i)$, we get for \eqref{eq:final_formula}
\begin{equation}
\begin{split}
    \tr{\P(\i)}&=\rho^{M-k}\left[\sum_{o_r\leq e}C_{e-o_r}C_{M-e+o_r-1}+\sum_{o_r> e}C_{o_r-e-1}C_{M-o_r+e}\right.\\
    &\qquad\qquad\left.+\rho^2\left(C_M-\sum_{o_r\leq e}C_{e-o_r}C_{M-e+o_r-1}-\sum_{o_r>e}C_{o_r-e-1}C_{M-o_r+e}\right)\right],
\end{split}
    \label{eq:formula_one_even}
\end{equation}
where $\P(\i)$ is cumbersome to write in this case, and $2o_r-1\notin\i$. See \tabref{tab:example_one_even} for some examples of the formula. In the case where $k=M$, i.e. there is only one odd number $2o-1$ missing from $\i$, we get a simpler formula
\begin{equation}
\tr{\P(\i)}=
\begin{cases}
\rho^{M-k}\left[C_{e-o}C_{M-e+o-1}+\rho^2\left(C_M-C_{e-o}C_{M-e+o-1}\right)\right],& o\leq e\\
\rho^{M-k}\left[C_{o-e-1}C_{M-o+e}+\rho^2\left(C_M-C_{o-e-1}C_{M-o+e}\right)\right],& o>e,
\end{cases}
\label{eq:formula_one_even_max_k}
\end{equation}
and 
\[
\P(\i)=
\begin{cases}
(\X\Xt)^{o-1}(\Xt)^2(\X\Xt)^{e-o-1}\X^2(\X\Xt)^{M-e},& o\leq e\\
(\X\Xt)^{e-1}(\X)^2(\X\Xt)^{o-e-1}(\Xt)^2(\X\Xt)^{M-o},& o>e.
\end{cases}\]
Furthermore, we see that in the case where $k=M$, i.e. there are as many matrices $\X$ as matrices $\Xt$, the polynomial from \eqref{eq:formula_one_even} has a constant term which corresponds to the moments of Ginibre matrices whenever $\rho=0$
\begin{equation}
\tr{\P(\i)}_{Gin.}=
\begin{cases}
C_{e-o}C_{M-e+o-1},& o\leq e\\
C_{o-e-1}C_{M-o+e},& o>e.
\end{cases}
\end{equation}

\begin{table}[t]
    \centering
    \begin{tabular}{cllcll}
    \toprule
        $M$ & \makecell[c]{$\i$} & \makecell[c]{$\{o_r\}$}  &  $e$  &  \makecell[c]{$\P(\i)$}  &  \makecell[c]{$\varphi(\P(\i))$}\\\hline
         $6$ & $(1,2,7,9)$ & $\{1,2,6\}$ & $1$ & $\X^2(\Xt)^4(\X\Xt)^2(\Xt)^2$ & $\rho^{2}(70+62\rho^2)$\\
         $6$ & $(3,6,7,9,11)$ & $\{1,3\}$ & $3$ & $(\Xt)^2\X(\Xt)^2\X^2\Xt(\X\Xt)^2$ & $\rho(52+80\rho^2)$\\
         $8$ & $(1,5,7,8,9,13)$ & $\{2,6,8\}$ & $4$ & $\X(\Xt)^3\X\Xt\X^3(\Xt)^3\X(\Xt)^3$ & $\rho^2(286+1144\rho^2)$\\
         $8$ & $(1,7,9,13,16)$ & $\{2,3,6,8\}$ & $8$ & $\X(\Xt)^5(\X\Xt)^2\X^2\Xt\X^2\Xt$ & $\rho^3(729+701\rho^2)$\\
    \botrule
    \end{tabular}
    \caption{Examples of \eqref{eq:formula_one_even} for different $M$ and indices tuples $\i$. The formula coincides with a numerical evaluation using SageMath \cite{sagemath}.}
    \label{tab:example_one_even}
\end{table}

\subsection{\texorpdfstring{$r=2$: $\i$ has exactly two even elements}{}}

Let us consider the case $k=M$. We denote by $2e_1<2e_2$, the two even elements in $\i$, and by $2o_1-1<2o_2-1$ the missing odd elements from $\i$. In this case, $\sigma_\circ(\pi)$ can take one of three values 0, 1, or 2. We can therefore write 
\[\tr{\P(\i)}=N_0(\i)+\rho N_1(\i)+\rho^2N_2(\i),\]
with $N_2(\i)=N_{\geq 2}(\i)$. We can start by expressing $N_2(\i)$ which yields
\begin{align*}
N_2(\i)&=\card{\Att_{e_1}(\i)\cap\Att_{e_2}(\i)}\\
&=C_M-\card{\Ac_{e_1,o_1}}-\card{\Ac_{e_1,o_2}}-\card{\Ac_{e_2,o_1}}-\card{\Ac_{e_2,o_2}}+\card{\Ac_{e_1,o_1}\cap\Ac_{e_2,o_2}}+\card{\Ac_{e_1,o_2}\cap\Ac_{e_2,o_1}}.
\end{align*}
To compute the cardinality of both $\Ac_{e_1,o_1}\cap\Ac_{e_2,o_2}$ and $\Ac_{e_1,o_2}\cap\Ac_{e_2,o_1}$, we need to consider the possible cases regarding the relative position between $o_1$, $o_2$, $e_1$ and $e_2$. Indeed, in the following let us consider for instance $e_1<o_1<o_2\leq e_2$ (see Appendix \ref{ap:computa_sec4} for the other cases). In this case, $\Ac_{e_1,o_2}\cap\Ac_{e_2,o_1}=\emptyset$ since if a pairing $\pi\in\nc_2(2M)$ had both edges $(2e_1,2o_2-1)$ and $(2o_1-1,2e_2)$, it would be crossing. On the other hand, it is possible to simultaneously form both edges $(2e_1,2o_1-1)$ and $(2e_2,2o_2-1)$ and 
\begin{equation}
\Ac_{e_1,o_1}\cap\Ac_{e_2,o_2}\cong \nc_2(2(o_1-e_1-1))\times\nc_2(2(e_2-o_2))\times\nc_2(2(M-e_2+o_2+e_1-o_1-1)).
\end{equation}
Consequently, $N_2(\i)$ can be expressed as
\begin{equation}
\begin{split}
N_2(\i)&=C_M-C_{o_1-e_1-1}C_{M-o_1+e_1}-C_{o_2-e_1-1}C_{M-o_2+e_1}-C_{e_2-o_1}C_{M-e_2+o_1-1}-C_{o_2-e_2-1}C_{M-o_2+e_2}\\
&\qquad +C_{o_1-e_1-1}C_{e_2-o_2}C_{M-e_2+o_2+e_1-o_1-1}.
\end{split}
\end{equation}
Now, we can use this result to compute $N_1(\i)$. Indeed,
\begin{align*}
N_1(\i)&=N_{\geq 1}(\i)-N_{\geq 2}(\i)\\
&=N_{\geq 1}(\i)-N_{2}(\i)\\
&=\card{\Att_{e_1}(\i)}+\card{\Att_{e_2}(\i)}-\card{\Att_{e_1}(\i)\cap \Att_{e_2}(\i)}-N_2(\i)\\
&=\card{\Att_{e_1}(\i)}+\card{\Att_{e_2}(\i)}-2N_2(\i)\\
&=C_{o_1-e_1-1}C_{M-o_1+e_1}+C_{o_2-e_1-1}C_{M-o_2+e_1}+C_{e_2-o_1}C_{M-e_2+o_1-1}+C_{o_2-e_2-1}C_{M-o_2+e_2}\\
&\qquad-2C_{o_1-e_1-1}C_{e_2-o_2}C_{M-e_2+o_2-e_1+o_1-1}
\end{align*}
Finally, $N_0(\i)$ can be obtained easily by using both expression for $N_1(\i)$ and $N_2(\i)$ such that
\begin{equation}
\begin{split}
\tr{\P(\i)}&=C_{o_1-e_1-1}C_{e_2-o_2}C_{M-e_2+o_2+e_1-o_1-1}\\
&+\rho^2\left(C_{o_1-e_1-1}C_{M-o_1+e_1}+C_{o_2-e_1-1}C_{M-o_2+e_1}+C_{e_2-o_1}C_{M-e_2+o_1-1}+C_{e_2-o_2}C_{M-e_2+o_2-1}\right.\\
&\left.\qquad-2C_{o_1-e_1-1}C_{e_2-o_2}C_{M-e_2+o_2+e_1-o_1-1}\right)\\
&+\rho^{4}\left(C_M-C_{o_1-e_1-1}C_{M-o_1+e_1}-C_{o_2-e_1-1}C_{M-o_2+e_1}-C_{e_2-o_1}C_{M-e_2+o_1-1}-C_{e_2-o_2}C_{M-e_2+o_2-1}\right.\\
&\left.\qquad +C_{o_1-e_1-1}C_{e_2-o_2}C_{M-e_2+o_2+e_1-o_1-1}\right),
\end{split}
\label{eq:final_r_2}
\end{equation}
where 
\[\P(\i)=(\X\Xt)^{e_1-1}\X^2(\X\Xt)^{o_1-e_1-1}(\Xt)^2(\X\Xt)^{o_2-o_1-1}(\Xt)^2(\X\Xt)^{e_2-o_2-1}\X^2(\X\Xt)^{M-e_2}.\]

\begin{table}[b]
    \centering
    \begin{tabular}{cllcll}
    \toprule
        $M$ & \makecell[c]{$\i$} & \makecell[c]{$\{o_r\}$}  &  $\{e_1,e_2\}$  &  \makecell[c]{$\P(\i)$}  &  \makecell[c]{$\varphi(\P(\i))$}\\\hline
        $8$ & $(2,3,9,10)$ & $\{1,3,4,6,7,8\}$ & $\{1,5\}$ & $\Xt\X^2(\Xt)^5\X^2(\Xt)^6$ & $\rho^4\left(564+734\rho^2+132\rho^4\right)$\\
        $8$ & $(1,3,5,6,9,10,13,15)$ & $\{4,6\}$ & $\{3, 5\}$ & $(\X\Xt)^2\X^2(\Xt)^2\X^2(\Xt)^2(\X\Xt)^2$ & $174+276\rho^2+530\rho^4$\\
        $8$ & $(5,7,9,11,13,14,15,16)$ & $\{1,2\}$ & $\{7,8\}$ & $(\Xt)^4(\X\Xt)^4\X^4$ & $42+693\rho^2+695\rho^4$\\
        $9$ & $(1,5,7,8,9,13,14,15)$ & $\{2,6,9\}$ & $\{4,7\}$ & $\X(\Xt)^3\X\Xt\X^3(\Xt)^3\X^3(\Xt)^3$ & $\rho\left(151+1655\rho^2+3056\rho^4\right)$\\
    \botrule
    \end{tabular}
    \caption{Examples of \eqref{eq:final_r_2} (and Appendix \ref{ap:computa_sec4}) for different $M$ and indices tuples $\i$. The formula coincides with a numerical evaluation using SageMath \cite{sagemath}.}
    \label{tab:example_two_even}
\end{table}

As an example, consider the case where $M=8$, $e_1=1$, $e_2=6$ and $o_1=3$, $o_2=4$ such that $\P(\i)=\X^2(\Xt)^5\X\Xt\X^3\Xt\X\Xt$. Applying the previous formula, we get
\[\tr{\X^2(\Xt)^5\X\Xt\X^3\Xt\X\Xt}=10+350\rho^2+1070\rho^4,\]
supported by numerical evaluation using SageMath \cite{sagemath}. See Table \ref{tab:example_two_even} for other examples. Furthermore, the constant term corresponds once again to the mixed-moments of Ginibre matrices such that
\[\tr{\P(\i)}_{Gin.}=C_{o_1-e_1-1}C_{e_2-o_2}C_{M-e_2+o_2+e_1-o_1-1},\]
in the present case where $e_1<o_1<o_2\leq e_2$. Taking $M=6$, $e_1=1$, $e_2=4$ and $o_1=3$, $o_2=6$, our formula gives (see Appendix \ref{ap:computa_sec4} and the corresponding case)
\[\tr{(\X^3(\Xt)^3)^2}_{Gin.}=4=FC_2(3),\]
as it should be. Finally, the general case where $k\leq M$ yields no particular difficulty aside from the lengthiness of the computations. Indeed, upon computing $\card{\Att_{e_1}(\i)\cap\Att_{e_2}(\i)}$ for $N_2(\i)$ one must consider all possible intersections $\Ac_{e_1,o_\ell}\cap\Ac_{e_2,o_s}$ where $2o_r-1,2o_\ell-1\notin \i$, see Appendix \ref{ap:computa_sec4} for a general formula.

\section{Conclusion and extension\label{sec:sec_final}}

In this article, we have found an explicit formula for the mixed-moments $\tr{\X^n(\Xt)^m}$ of Gaussian Elliptic Matrices $\X$ with a correlation parameter $\rho$ between elements $ij$ and $ji$. Using both combinatorial methods and mapping onto Temperley-Lieb algebras, we showed that these mixed-moments could be expressed as a polynomial in $\rho$ whose coefficients enumerate specific non-crossing partitions (or equivalently specific Temperley-Lieb diagrams) over $n+m$ elements, an enumeration performed by Catalan triangular numbers \cite{carlitz1972sequences}. This result fits in an already rich body of literature linking the computation of mixed-moments (especially of Gaussian-type matrices) to Catalan structures \cite{halmagyi_mixed_2020, kemp_enumeration_2011}. However, the explicit link with ballot numbers that we find in this article is novel, to the best of our knowledge. Furthermore, in our specific case, the gain in computing power from the general formula \eqref{eq:final_formula} is quite substantial since we can reduce the complexity from exponential (coming from the enumeration of non-crossing pairings) to polynomial. In the context of conewise linear dynamics from \cite{Dessertaine2022Cones}, this result shows that, in the large-dimensional limit, the correlator of the Gaussian process has the form
\[\avgoe{v_i(t)v_j(s)}\underset{t,s\to\infty}{\sim}\delta_{ij}\Psi_{\frac{t}{s}}(\rho)e^{-(t+s)\hat{\Phi}_{\frac{t}{s}}(\rho)},\]
assuming $t\leq s$. In the general case, one cannot link directly this asymptotic behavior to the decay of the persistence of the process. However, if $t+s\ll \hat{\Phi}_{\frac{t}{s}}(\rho)$, the correlator is approximately self-similar, such that the persistence decays algebraically with an exponent $\mu(\rho)$ at small-times (see \cite{NewellRosenblatt1962,BrayReviewPersistence}). 

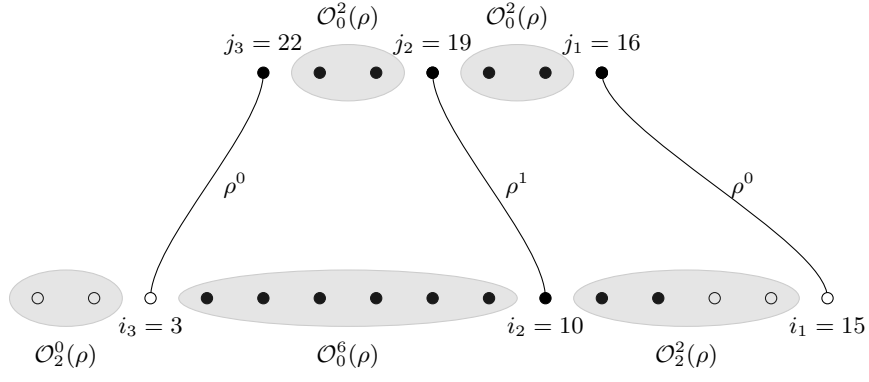
\begin{figure}
    \centering
    \begin{tikzpicture}[scale=0.75, every node/.style={scale=1}]
    \draw (-1,0) [dnUp={2}{white}{black}];
    \draw (6,0) [dnUp={-2}{black}{black}];
    \draw (11,0) [dnUp={-4}{white}{black}];
    
    \draw (0,0) node[ptm={black}] {} {};
    \draw (-2,0) node[ptm={white}] {};
    \draw (-3,0) node[ptm={white}] {};
    \draw (1,0) node[ptm={black}] {};
    \draw (2,0) node[ptm={black}] {};
    \draw (3,0) node[ptm={black}] {};
    \draw (4,0) node[ptm={black}] {};
    \draw (5,0) node[ptm={black}] {};
    \draw (7,0) node[ptm={black}] {};
    \draw (8,0) node[ptm={black}] {};
    \draw (9,0) node[ptm={white}] {};
    \draw (10,0) node[ptm={white}] {};
    
    \draw (7,4) node[ptm={black}] {};
    \draw (6,4) node[ptm={black}] {};
    %\draw (6,4) node[ptm={black}] {};
    %\draw (7,4) node[ptm={black}] {};
    %\draw (8,4) node[ptm={black}] {};
    
    \draw (2,4) node[ptm={black}] {};
    \draw (3,4) node[ptm={black}] {};
    \draw (4,4) node[ptm={black}] {};
    \draw (5,4) node[ptm={black}] {};
    %\draw (1,0) node[ptm={white}] {};

    \draw[fill=gray, opacity=0.2] (2.5,0) ellipse (3cm and 0.5cm);
    \draw[fill=gray, opacity=0.2] (-2.5,0) ellipse (1cm and 0.5cm);
    \draw[fill=gray, opacity=0.2] (8.5,0) ellipse (2cm and 0.5cm);
    \draw[fill=gray, opacity=0.2] (2.5,4) ellipse (1cm and 0.5cm);
    \draw[fill=gray, opacity=0.2] (5.5,4) ellipse (1cm and 0.5cm);
    
    \node at (2.5,5) {$\mm{0}{2}(\rho)$};
    %\node at (1.5,5) {$\mm{0}{2}(\rho)$};
    \node at (5.5,5) {$\mm{0}{2}(\rho)$};
    \node at (-2.5,-1) {$\mm{2}{0}(\rho)$};
    \node at (2.5,-1) {$\mm{0}{6}(\rho)$};
    \node at (8.5,-1) {$\mm{2}{2}(\rho)$};
    
    \node at (-1,-0.5) {$i_3=3$};
    %\node at (1.5,5) {$p^{(1)}_{0,2}(\rho)$};
    \node at (1,4.5) {$j_3=22$};
    \node at (6,-0.5) {$i_2=10$};
    \node at (4,4.5) {$j_2=19$};
    \node at (11,-0.5) {$i_1=15$};
    \node at (7,4.5) {$j_1=16$};
    
    \node at (0.5,2) {$\rho^0$};
    \node at (5.5,2) {$\rho^1$};
    \node at (9.5,2) {$\rho^0$};
\end{tikzpicture}
    \caption{Example of a specific term involved in the recursive formula to compute $\tr{\X^3(\Xt)^9\X^3(\Xt)^7}$.}
    \label{fig:conclu_rec}
\end{figure}

The enumeration carried out in \secref{sec:sec2} could very well be adapted to compute higher order mixed-moments $\mm{\n}{\m}(\rho)$. Indeed, considering for instance $\tr{\X^{n_1}(\Xt)^{m_1}\X^{n_2}(\Xt)^{m_2}}$, one could consider that $\X^{n_1}(\Xt)^{m_1}\X^{n_2}(\Xt)^{m_2}$ is formed of two "blocks" $\X^{n_1}(\Xt)^{m_1}\X^{n_2}$ and $(\Xt)^{m_2}$ and partition $\nc_2(n_1+n_2+m_1+m_2)$ into subsets with a constant number of mixed-blocks pairs, i.e. linking a matrix from the first block with the second. Depending on the position of the mixed pairs, the sub-pairings imposed by the non-crossing structure would contribute lower-order moments to the enumeration. For instance, consider the moment $\tr{\X^3(\Xt)^9\X^3(\Xt)^7}$ as in \figref{fig:conclu_rec}, and three mixed-blocks pairs, the sub-pairings would yield five mixed-moments of the type $\mm{n}{m}(\rho)$. Summing over all possible mixed-block pairs (and not forgetting the $\rho$-contribution of each one, e.g. a $\rho$ contribution for $(i_2,j_2)$ in \figref{fig:conclu_rec}), we would get a recursive formula for higher order mixed-moments.

In \secref{sec:sec4}, we then provided three explicit formulas for specific cases of general mixed-moments of \gee. Instead of grouping matrices $\X$ and $\Xt$ into blocks of size $n_i$ and $m_i$ as in \cite{kemp_enumeration_2011,halmagyi_mixed_2020}, we encoded the positions of matrices $\X$ into an ordered tuple $\i$. Given $\i$, we provided in \eqref{eq:general_gem} and \eqref{eq:N_geq_j} a general abstract formula for the moments of \gee. In the specific cases where $\i$ has $0$, $1$, or $2$ even elements, we were able to explicitly carry out the computations to give closed-form formulas \eqref{eq:final_r_0}, \eqref{eq:formula_one_even}, \eqref{eq:final_r_2} (see also Appendix \ref{ap:computa_sec4}) for the mixed-moments. Once again, these explicit formulas are computationally efficient since one need not enumerate all pairings in $\nc_2(2M)$. Finally, it could be possible to further investigate the general formula \eqref{eq:N_geq_j} since the innermost quantity can be rewritten as $\bigcap_{u=1}^s{\Att_{\k_u}(\i)}=\Att_{\bigcup_{u=1}^s\k_u}(\i)$, allowing to group together sets sharing the same $\bigcup_{u=1}^s\k_u$. We look forward to working in this direction in the near future.

\subsection*{Acknowledgments}
I want to thank Pierre Mergny for illuminating discussions on Random Matrix Theory. I also want to warmly thank Jean-Philippe Bouchaud for his comments and remarks on the paper. This research was partially conducted within the Econophysics \& Complex Systems Research Chair, under the aegis of the Fondation du Risque, the Fondation de l’Ecole polytechnique, the
Ecole polytechnique and Capital Fund Management.

\appendix
% !TEX root = ./main.tex
\setcounter{table}{0}
\renewcommand{\thetable}{A\arabic{table}}

\section{\texorpdfstring{Catalan triangular numbers and cardinality of $\nc_2^{2k}(2u,2v)$}{}\label{ap:catalan_triangle}}

Catalan triangular numbers $C(n,k)$ are a generalization of Catalan numbers defined through the relation
\begin{equation}
    C(n,k)=C(n,k-1)+C(n-1,k),
\end{equation}
with $C(n,1)=1$ and $C(n,k)=0$ for $k>n$. Using combinatorics, see \cite{carlitz1972sequences} for instance, one can show that $C(n,k)$ has the exact expression
\begin{equation}
    C(n,k)=\frac{n-k-1}{n}\binom{n+k-2}{n-1}.
\end{equation}
Note that $C(n+1,n+1)=C_{n}$, as shown in \ref{tab:catalan}. Furthermore, in the language of \cite{AIGNER199933}, one can show that Catalan triangular numbers form an \emph{admissible sequence} which satisfies
\begin{equation}
    C(a+b,b)=\sum_{s=1}^b C(s,s)C(a+b-s,b-a+1).
    \label{eq:properties_catalan_triangle}
\end{equation}
From this property, one can show that the numbers $B(k,t)$ from \eqref{eq:def_B} in the main text, can be expressed as
\begin{equation}
    B(k,t)=C(t+k+1,t-k+1).
\end{equation}
Indeed, using \eqref{eq:rec_B} and assuming that the previous property holds for all $q<k$ and $s<t$, we have the following computation
\begin{align*}
    B(k,t)&=\sum_{r=k}^{t}C_{t-r}\sum_{s=k}^{r}C_{r-s}B(k-1,s-1)\\
    &=\sum_{r=k}^{t}C(t-r+1,t-r+1)\sum_{s=k}^{r}C(r-s+1,r-s+1)C(s-1+k,s-1-k+2)\\
    &=\sum_{r=k}^{t}C(t-r+1,t-r+1)\sum_{j=1}^{r-k+1}C(j,j)C(r+k-j,r-k-j+2)\\
    &=\sum_{r=k}^{t}C(t-r+1,t-r+1)C(r+k,r-k+1)\\
    &=\sum_{j=1}^{t-k+1}C(j,j)C(t-j+k+1,t-j-k+2)\\
    &=C(t+k+1,t-k+1).
\end{align*}
This proves our result by induction since base cases are straightforwardly satisfied.

\begin{table}[b]
\fontsize{10}{12}\selectfont
    \centering
    \begin{tabular}{|p{1cm}|*{7}{p{1cm}|}}\hline
\backslashbox{$n$}{$k$}
& 1& 2 & 3
& 4 & 5 & 6 & 7\\\hline
1 & 1 & 0 & 0 & 0 & 0 & 0 & 0\\\hline
2 & 1 & 1 & 0 & 0 & 0 & 0 & 0\\\hline
3 & 1 & 2 & 2 & 0 & 0 & 0 & 0\\\hline
4 & 1 & 3 & 5 & 5 & 0 & 0 & 0\\\hline
5 & 1 & 4 & 9 & 14& 14 & 0 & 0\\\hline
6 & 1 & 5 & 14 & 28 & 42 & 42 & 0\\\hline
7 & 1 & 6 & 20 & 48 & 90 & 132 & 132\\\hline
\end{tabular}
    \caption{Catalan triangle containing the numbers $C(n,k)$}
    \label{tab:catalan}
\end{table}

\section{\texorpdfstring{Detailed computation for the asymptotic behavior of $\mm{n}{m}(\rho)$}{}\label{ap:asymptotic}}

As in the main text, we will consider the case where $n=2u$ and $m=2v$ are even. Without loss of generality (using the invariance of $\tr{\cdot}$ under transposition), we will consider $n\geq m$ and we will denote by $q=n/m=u/v\geq 1$. We will assume that $q$ is kept constant as $n,m\to\infty$. We recall here the formula from the main text
\begin{equation}
    \mm{2u}{2v}(\rho)=\rho^{u+v}\sum_{k=0}^v\rho^{-2k}\frac{(2k+1)^2}{(u+k+1)(v+k+1)}\binom{2u}{u+k}\binom{2v}{v+k}.
\end{equation}
Let us introduce an auxiliary quantity $Q_{u,v}(x)$ defined for $x\geq 1$ as
\begin{equation}
    Q_{u,v}(x)=\sum_{k=0}^v\frac{x^{2k}}{(u+k+1)(v+k+1)}\binom{2u}{u+k}\binom{2v}{v+k},
\end{equation}
along with the differential operator $D_x(\cdot)=\frac{\d}{\d x}\left(-x\left(\cdot\right)+\frac{\d}{\d x}\left(x^{2}\left(\cdot\right)\right)\right)$. A straightforward computation shows that 
\begin{equation}
    \mm{2u}{2v}(\rho)=\rho^{u+v}\left.D_xQ_{u,v}\right|_{x=\rho^{-1}}
    \label{eq:rel_P_Q}
\end{equation}

\subsection{\texorpdfstring{A saddle point approximation for $Q_{u,v}(x)$}{}}

As $u,v\to\infty$ the binomial terms of the sum defining $Q_{u,v}(x)$ can be approximated as follows
\begin{align}
    \binom{2v}{v+k}&\sim \frac{4^v}{\sqrt{\pi v\left(1-(k/v)^2\right)}}\left[\left(1-\frac{k}{v}\right)^{1-\frac{k}{v}}\right]^{-v}\left[\left(1+\frac{k}{v}\right)^{1+\frac{k}{v}}\right]^{-v}\\
    \binom{2u}{u+k}&\sim \frac{4^{qv}}{\sqrt{q\pi v\left(1-(k/(qv))^2\right)}}\left[\left(1-\frac{k}{qv}\right)^{1-\frac{k}{qv}}\right]^{-qv}\left[\left(1+\frac{k}{qv}\right)^{1+\frac{k}{qv}}\right]^{-qv}.
\end{align}
Injecting these expressions back into $Q_{u,v}$, we see that it can be expressed as a Riemann sum
\[Q_{u,v}(x)\underset{\substack{u,v\to\infty\\u/v=q}}{\sim} \frac{4^{q(v+1)}}{\pi q^{3/2} v^3}\sum_{k=0}^vg_q(k/v)e^{v\Fc_q(x,k/v)},\]
with
\begin{align*}
    g_q(y)&=\left((1+y)(1+y/q)\sqrt{(1-y^2)\left(1-\frac{y^2}{q^2}\right)}\right)^{-1},\\
    \Fc_q(x,y)&=y\log{x^2}-(1+y)\log{(1+y)}-(1-y)\log{(1-y)}\\
    &\qquad-\left(1+\frac{y}{q}\right)\log{\left(1+\frac{y}{q}\right)}-\left(1-\frac{y}{q}\right)\log{\left(1-\frac{y}{q}\right)}.
\end{align*}
We can therefore approximate $Q_{u,v}(x)$ by the following integral
\begin{equation}
    Q_{u,v}(x)\underset{\substack{u,v\to\infty\\u/v=q}}{\sim}\frac{4^{q(v+1)}}{\pi q^{3/2} v^2}\int_0^1\d y\, g_q(y)e^{v\Fc_q(x,y)}.
\end{equation}
With our assumption $q\geq 1$, i.e. $u\geq v$, and the fact that $x\geq 1$, the function $\Fc_q(x,y)$ reaches a global maximum on $[0,1]$ at a point $\sd{q}(x)$ solution of $\partial\Fc_q(x,y)/\partial y=0$ which reads
\begin{equation}
    \sd{q}(x)=\frac{(q+1)(x^2+1)-\sqrt{(q+1)^2(x^2+1)^2-4q(x^2-1)^2}}{2(x^2-1)}.
\end{equation}
Calling $\mathfrak{g}$, the generating function of the Catalan numbers, i.e.
\[\mathfrak{g}(z)=\sum_{n=0}^{\infty}z^{n}C_{n}=\frac{1-\sqrt{1-4z}}{2z},\]
the saddle point can be written as
\begin{equation}
    \sd{q}(x)=\frac{\sqrt{q}}{q+1}\frac{x^2-1}{x^2+1}\,\mathfrak{g}\left(\frac{q}{(q+1)^2}\left(\frac{x^2-1}{x^2+1}\right)^2\right).
\end{equation}
Interestingly, the factor $q/(q+1)^2=uv/(u+v)^2$ is related to the asymptotic behavior of the Catalan numbers since
\[\frac{C_{u+v}}{\sqrt{C_{2u}C_{2v}}}\sim\left(\frac{4uv}{(u+v)^2}\right)^3.\]
Using a saddle point approximation, we can give a final asymptotic estimate for $Q_{u,v}(x)$
\begin{equation}
    Q_{u,v}(x)\underset{\substack{u,v\to\infty\\u/v=q}}{\sim}\sqrt{\frac{2}{\pi}}\frac{4^{q(v+1)}}{q^{3/2} v^{5/2}}\left(-\frac{\partial^2\Fc_q}{\partial y^2}(x,\sd{q}(x))\right)^{-1/2}g_q(\sd{q}(x))e^{v\Fc_q(x,\sd{q}(x))}.
    \label{eq:asymp_Q}
\end{equation}
In the following, we will denote by $H_q$ the function
\[H_q(y)=\left(-\frac{\partial^2\Fc_q}{\partial y^2}(x,y)\right)^{-1/2}g_q(y)=\sqrt{\frac{q}{2(q+1)}}\left[(1+y)\left(1+\frac{y}{q}\right)\sqrt{1-\frac{y^2}{q}}\right]^{-1}.\]

\subsection{\texorpdfstring{Asymptotic behavior for $\mm{2u}{2v}(\rho)$}{}}

Combining \eqref{eq:rel_P_Q} with \eqref{eq:asymp_Q}, we get 
\begin{equation}
    \mm{2u}{2v}(\rho)\underset{\substack{u,v\to\infty\\u/v=q}}{\sim}\rho^{u+v}\sqrt{\frac{2}{\pi}}\frac{4^{q(v+1)}}{q^{3/2} v^{5/2}}\left.R(v,x)e^{v\Fc_q(x,\sd{q}(x))}\right|_{x=\rho^{-1}},
    \label{eq:general_behavior}
\end{equation}
where $R(v,x)$ is a second order polynomial in $v$ with $x$-dependent coefficients $(r_i(x))_{i=0,1,2}$ such that
\begin{equation}
    \begin{aligned}
    r_0(x)&=H_q(\sd{q}(x))+3x\frac{\d\sd{q}}{\d x}\frac{\d H_{q}}{\d x}(\sd{q}(x))+x^2\frac{\d^2\sd{q}}{\d x^2}\frac{\d H_{q}}{\d x}(\sd{q}(x))+x^2\left(\frac{\d\sd{q}}{\d x}\right)^2\frac{\d^2H_{q}}{\d x^2}(\sd{q}(x))\\
    r_1(x)&=4\sd{q}(x)H_q(\sd{q}(x))+4x\sd{q}(x)\frac{\d\sd{q}}{\d x}\frac{\d H_{q}}{\d x}(\sd{q}(x))+x\frac{\d\sd{q}}{\d x}H_{q}(\sd{q}(x))\left(4+x\frac{\d\sd{q}}{\d x}\frac{\partial^2 \Fc_q}{\partial y^2}(x,\sd{q}(x)\right)\\
    r_2(x)&=4(\sd{q}(x))^2H_q(\sd{q}(x)),
    \end{aligned}
\end{equation}
where we used the relationships
\begin{align*}
    &\frac{\partial\Fc_q}{\partial x}(x,y)=2\frac{y}{x}\;,\;\frac{\partial^2\Fc_q}{\partial x^2}(x,y)=-2\frac{y}{x^2}\\
    &\frac{\partial\Fc_q}{\partial y}(x,\sd{q}(x))=0\;,\; \frac{\partial^2\Fc_q}{\partial x\partial y}(x,y)=\frac{2}{x}.
\end{align*}
Now, one must be particularly careful of the interaction between the position of the saddle point $\sd{q}(x)$ with respect to the width of the Gaussian function used in the saddle point approximation $\left(-v\frac{\partial^2\Fc_q}{\partial y^2}(x,\sd{q}(x))\right)^{-1/2}$

\subsubsection{\texorpdfstring{Behavior for $\rho\to 1^-$}{}}

In the limit $\rho\to1^-$, we should be able to recover the asymptotic behavior of the $(u+v)-$th Catalan number since $\mm{2u}{2v}(1)=C_{u+v}$. In this limit, which is equivalent to $x\to1^+$ with the previous notations, the saddle point has the following behavior
\begin{equation}
    \sd{q}(x)\underset{x\to1^+}{\sim}\frac{q}{q+1}(x-1)\to0^+.
\end{equation}
This implies two regimes with the following implications:
\begin{enumerate}[label=$(\roman*)$]
    \item $\sd{q}(x)\gg\left(-v\frac{\partial^2\Fc_q}{\partial y^2}(x,\sd{q}(x))\right)^{-1/2}$, i.e. even though the saddle point goes towards zero it is still much larger than the Gaussian width: in this case, \eqref{eq:general_behavior} is valid and we are effectively in the case of \secref{sec:general_beahvior}.
    \item $\sd{q}(x)\ll\left(-v\frac{\partial^2\Fc_q}{\partial y^2}(x,\sd{q}(x))\right)^{-1/2}$, i.e. the saddle point goes towards zero and is much smaller than the Gaussian width: in this case, \eqref{eq:general_behavior} is not valid since the saddle point approximation must be performed on a \emph{half-Gaussian}.
\end{enumerate}

Placing ourselves in case $(ii)$, we must apply a factor $\frac{1}{2}$ to \eqref{eq:general_behavior} because of the integration on the half-Gaussian. Furthermore, since
\begin{subequations}
    \begin{align}
    \Fc_q(x,\sd{q}(x))&\sim\frac{q}{q+1}(\sd{q}(x))^2\\
    -\frac{\partial^2\Fc_q}{\partial y^2}(x,\sd{q}(x))&\longrightarrow 2\frac{q+1}{q}\\
    H_q(\sd{q}(x))&\longrightarrow\sqrt{\frac{q}{2(q+1)}}\\
    \frac{\d H_q}{\d x}(\sd{q}(x))&\longrightarrow-\sqrt{\frac{q+1}{2q}}\\
    \frac{\d^2 H_q}{\d x^2}(\sd{q}(x))&\longrightarrow\sqrt{\frac{q}{2(q+1)}}\frac{2+3q+2q^2}{q^2}\\
    \frac{\d \sd{q}}{\d x}(x)&\longrightarrow\frac{q}{q+1}\\
    \frac{\d^2 \sd{q}}{\d x^2}(x)&\longrightarrow-\frac{q}{q+1}
    \end{align}
\end{subequations}
and therefore $v^2(\sd{q}(x))^2\ll v$,
the polynomial $R(v,x)$ from \eqref{eq:general_behavior} is dominated by $vr_1(x)$. Using the previous limits, we get that \begin{equation*}
    vr_1(x)\sim vx\frac{\d\sd{q}}{\d x}H_{q}(\sd{q}(x))\left(4+x\frac{\d\sd{q}}{\d x}\frac{\partial^2 \Fc_q}{\partial y^2}(x,\sd{q}(x)\right)\sim v\sqrt{2}\left(\frac{q}{q+1}\right)^{3/2}.
\end{equation*}
Furthermore, the argument $v\Fc_q(x,\sd{q}x))$ of the exponential term is such that $v\Fc_q(x,\sd{q}x))\sim v\frac{q}{q+1}(\sd{q}(x))^2\ll 1$, so that the exponential does not contribute in the present case. Finally, injecting all previous results into \eqref{eq:general_behavior} (and remembering to apply the $\frac{1}{2}$ factor), we get
\begin{equation}
    \mm{2u}{2v}(\rho\to1^-)\underset{\substack{u,v\to\infty\\u/v=q}}{\sim}\frac{4^{q(v+1)}}{\sqrt{\pi}v^{3/2}(q+1)^{3/2}}=\frac{4^{u+v}}{\sqrt{\pi}(u+v)^{3/2}}\underset{u,v\to\infty}{\sim} C_{u+v}.,
\end{equation}
and retrieve the asymptotic behavior of the Catalan numbers.

\subsubsection{\texorpdfstring{Behavior for generic $\rho$}{}\label{sec:general_beahvior}}

Whenever $\sd{q}(x)\gg\left(-v\frac{\partial^2\Fc_q}{\partial y^2}(x,\sd{q}(x))\right)^{-1/2}$, the saddle point approximation \eqref{eq:general_behavior} holds and the dominating terms in $R(v,x)$ reverse such that $R(v,x)\sim v^2r_2(x)$. Consequently, \eqref{eq:general_behavior} can be rewritten as
\begin{equation}
\mm{2u}{2v}(\rho)\underset{\substack{u,v\to\infty\\u/v=q}}{\sim}\rho^{u+v}\sqrt{\frac{2}{\pi}}\frac{4^{u+v}}{q^{3/2} v^{1/2}}\left.r_2(x)e^{v\Fc_q(x,\sd{q}(x))}\right|_{x=\rho^{-1}}.
\end{equation}
Since $\sd{q}(1/\rho)=-\sd{q}(\rho)$ and $\Fc_q(1/\rho,\sd{q}(1/\rho))=\Fc_q(\rho,\sd{q}(\rho))$, we finally get 
\begin{equation}
\mm{2u}{2v}(\rho)\underset{\substack{u,v\to\infty\\u/v=q}}{\sim}4\sqrt{\frac{2}{\pi}}\frac{(4\rho)^{u+v}}{q^{3/2} v^{1/2}}H_q\left(-\sd{q}(\rho)\right)(\sd{q}(\rho))^2e^{v\Fc_q(\rho,\sd{q}(\rho))}.
\end{equation}

Considering $\rmm{2u}{2v}$ as in the main text, we get 
\begin{equation}
    \rmm{2u}{2v}(\rho)\underset{\substack{u,v\to\infty\\u/v=q}}{\sim}\Psi_q(\rho)e^{-v\Phi_q(\rho)},
\end{equation}
with 
\begin{align}
    \Psi_q(\rho)&=16q^{-5/4}\sqrt{\rho}\frac{H_q(-\sd{q}(\rho))}{(1-\rho)^2(1+\rho)}\\
    \Phi_q(\rho)&=-\Fc_q(\rho,\sd{q}(\rho))+\frac{1}{2}(q+1)\Fc_1(\rho,\sd{1}(\rho)).
\end{align}

\section{\texorpdfstring{Cases for the computation of $N_2(\i)$ in the case where $\i$ has two even elements.}{}\label{ap:computa_sec4}}

\subsection{\texorpdfstring{Cardinality of $\Att_{e_1}\cap\Att_{e_2}$ and subsequent mixed-moment in the case $k=M$}{}}

\subsubsection{\texorpdfstring{Case $o_1<o_2\leq e_1<e_2$}{}}

In this case, $\Ac_{e_1,o_1}\cap\Ac_{e_2,o_2}=\emptyset$ since any pairing in this set would be crossing. On the other hand,
\[\Ac_{e_1,o_2}\cap\Ac_{e_2,o_1}\cong\nc_2(2(e_1-o_2))\times\nc_2(2(e_2-e_1+o_2-o_1-1))\times\nc_2(2(M-e_2+o_1-1)).\]
Consequently, we get the expression for the mixed-moment
\begin{equation}
\begin{split}
\tr{\P(\i)}&=C_{e_1-o_2}C_{e_2-e_1+o_2-o_1-1}C_{M-e_2+o_1-1}\\
&+\rho^2\left(C_{e_1-o_1}C_{M-e_1+o_1-1}+C_{e_1-o_2}C_{M-e_1+o_2-1}+C_{e_2-o_1}C_{M-e_2+o_1-1}+C_{e_2-o_2}C_{M-e_2-o_2-1}\right.\\
&\left.\qquad-2C_{e_1-o_2}C_{e_2-e_1+o_2-o_1-1}C_{M-e_2+o_1-1}\right)\\
&+\rho^{4}\left(C_M-C_{e_1-o_1}C_{M-e_1+o_1-1}-C_{e_1-o_2}C_{M-e_1+o_2-1}-C_{e_2-o_1}C_{M-e_2+o_1-1}-C_{e_2-o_2}C_{M-e_2-o_2-1}\right.\\
&\left.\qquad +C_{e_1-o_2}C_{e_2-e_1+o_2-o_1-1}C_{M-e_2+o_1-1}\right),
\end{split}
\end{equation}
with 
\begin{equation}
\P(\i)=(\X\Xt)^{o_1-1}(\Xt)^2(\X\Xt)^{o_2-o_1-1}(\Xt)^2(\X\Xt)^{e_1-o_2-1}\X^2(\X\Xt)^{e_2-e_1-1}\X^2(\X\Xt)^{M-e_2}.
\end{equation}

\subsubsection{\texorpdfstring{Case $o_1\leq e_1<o_2\leq e_2$}{}}

In this case, both $\Ac_{e_1,o_1}\cap\Ac_{e_2,o_2}$ and $\Ac_{e_1,o_2}\cap\Ac_{e_2,o_1}$ are non-empty and 
\[\begin{aligned}
\Ac_{e_1,o_1}\cap\Ac_{e_2,o_2}&\cong\nc_2(2(e_2-o_2))\times\nc_2(2(e_1-o_2))\times\nc_2(2(M-e_1-e_2+o_1+o_2-2))\\
\Ac_{e_1,o_2}\cap\Ac_{e_2,o_1}&\cong\nc_2(2(e_2-o_1-1))\times\nc_2(2(e_2+e_1-o_2-o_1))\times\nc_2(2(M-e_2+o_1-1)).
\end{aligned}\]
Consequently, we get the expression for the mixed-moment
\begin{equation}
\begin{split}
\tr{\P(\i)}&=C_{e_2-o_2}C_{e_1-o_1}C_{M-e_1-e_2+o_1+o_2-2}+C_{o_2-e_1-1}C_{e_2-o_2+e_1-o_1}C_{M-e_2+o_1-1}\\
&+\rho^2\left(C_{e_1-o_1}C_{M-e_1+o_1-1}+C_{o_2-e_1-1}C_{M-o_2+e_1}+C_{e_2-o_1}C_{M-e_2+o_1-1}+C_{e_2-o_2}C_{M-e_2+o_2-1}\right.\\
&\left.\qquad-2C_{e_2-o_2}C_{e_1-o_1}C_{M-e_1-e_2+o_1+o_2-2}-2C_{o_2-e_1-1}C_{e_2-o_2+e_1-o_1}C_{M-e_2+o_1-1}\right)\\
&+\rho^{4}\left(C_M-C_{e_1-o_1}C_{M-e_1+o_1-1}-C_{o_2-e_1-1}C_{M-o_2+e_1}-C_{e_2-o_1}C_{M-e_2+o_1-1}-C_{e_2-o_2}C_{M-e_2+o_2-1}\right.\\
&\left.\qquad +C_{e_2-o_2}C_{e_1-o_1}C_{M-e_1-e_2+o_1+o_2-2}+C_{o_2-e_1-1}C_{e_2-o_2+e_1-o_1}C_{M-e_2+o_1-1}\right),
\end{split}
\end{equation}
with 
\begin{equation}
\P(\i)=(\X\Xt)^{o_1-1}(\Xt)^2(\X\Xt)^{e_1-o_1-1}(\X)^2(\X\Xt)^{o_2-e_1-1}(\Xt)^2(\X\Xt)^{e_2-o_2-1}\X^2(\X\Xt)^{M-e_2}.
\end{equation}

\subsubsection{\texorpdfstring{Case $o_1\leq e_1<e_2< o_2$}{}}

In this case, $\Ac_{e_1,o_2}\cap\Ac_{e_2,o_1}=\emptyset$ and 
\[\Ac_{e_1,o_1}\cap\Ac_{e_2,o_2}\cong\nc_2(2(e_1-o_1))\times\nc_2(2(o_2-e_2-1))\times\nc_2(2(M-o_2+e_2-e_1+o_1-1)).\]
Consequently, we get the expression for the mixed-moment
\begin{equation}
\begin{split}
\tr{\P(\i)}&=C_{e_1-o_1}C_{o_2-e_2-1}C_{M-o_2+e_2-e_1+o_1-1}\\
&+\rho^2\left(C_{e_1-o_1}C_{M-e_1+o_1-1}+C_{o_2-e_1-1}C_{M-o_2+e_1}+C_{e_2-o_1}C_{M-e_2+o_1-1}+C_{o_2-e_2-1}C_{M-o_2+e_2}\right.\\
&\left.\qquad-2C_{e_1-o_1}C_{o_2-e_2-1}C_{M-o_2+e_2-e_1+o_1-1}\right)\\
&+\rho^{4}\left(C_M-C_{e_1-o_1}C_{M-e_1+o_1-1}-C_{o_2-e_1-1}C_{M-o_2+e_1}-C_{e_2-o_1}C_{M-e_2+o_1-1}-C_{o_2-e_2-1}C_{M-o_2+e_2}\right.\\
&\left.\qquad +C_{e_1-o_1}C_{o_2-e_2-1}C_{M-o_2+e_2-e_1+o_1-1}\right),
\end{split}
\end{equation}
with 
\begin{equation}
\P(\i)=(\X\Xt)^{o_1-1}(\Xt)^2(\X\Xt)^{e_1-o_1-1}\X^2(\X\Xt)^{e_2-e_1-1}\X^2(\X\Xt)^{o_2-e_2-1}(\Xt)^2(\X\Xt)^{M-o_2}.
\end{equation}

\subsubsection{\texorpdfstring{Case $e_1<o_1<o_2\leq e_2$}{}}

In this case, $\Ac_{e_1,o_2}\cap\Ac_{e_2,o_1}=\emptyset$ and 
\[\Ac_{e_1,o_1}\cap\Ac_{e_2,o_2}\cong\nc_2(2(o_1-e_1-1))\times\nc_2(2(e_2-o_2))\times\nc_2(2(M-e_2+o_2+e_1-o_1-1)).\]
Consequently, we get the expression for the mixed-moment
\begin{equation}
\begin{split}
\tr{\P(\i)}&=C_{o_1-e_1-1}C_{e_2-o_2}C_{M-e_2+o_2+e_1-o_1-1}\\
&+\rho^2\left(C_{o_1-e_1-1}C_{M-o_1+e_1}+C_{o_2-e_1-1}C_{M-o_2+e_1}+C_{e_2-o_1}C_{M-e_2+o_1-1}+C_{e_2-o_2}C_{M-e_2+o_2-1}\right.\\
&\left.\qquad-2C_{o_1-e_1-1}C_{e_2-o_2}C_{M-e_2+o_2+e_1-o_1-1}\right)\\
&+\rho^{4}\left(C_M-C_{o_1-e_1-1}C_{M-o_1+e_1}-C_{o_2-e_1-1}C_{M-o_2+e_1}-C_{e_2-o_1}C_{M-e_2+o_1-1}-C_{e_2-o_2}C_{M-e_2+o_2-1}\right.\\
&\left.\qquad +C_{o_1-e_1-1}C_{e_2-o_2}C_{M-e_2+o_2+e_1-o_1-1}\right),
\end{split}
\end{equation}
with
\[\P(\i)=(\X\Xt)^{e_1-1}\X^2(\X\Xt)^{o_1-e_1-1}(\Xt)^2(\X\Xt)^{o_2-o_1-1}(\Xt)^2(\X\Xt)^{e_2-o_2-1}\X^2(\X\Xt)^{M-e_2}.\]

\subsubsection{\texorpdfstring{Case $e_1<o_1\leq e_2<o_2$}{}}

In this case, both $\Ac_{e_1,o_1}\cap\Ac_{e_2,o_2}$ and $\Ac_{e_1,o_2}\cap\Ac_{e_2,o_1}$ are non-empty and 
\[\begin{aligned}
\Ac_{e_1,o_1}\cap\Ac_{e_2,o_2}&\cong\nc_2(2(o_1-e_1-1))\times\nc_2(2(o_2-e_2-1))\times\nc_2(2(M-o_2-o_1+e_1+e_2))\\
\Ac_{e_1,o_2}\cap\Ac_{e_2,o_1}&\cong\nc_2(2(e_2-o_1))\times\nc_2(2(o_2+o_1-e_2-e_1-2))\times\nc_2(2(M-o_2+e_1)).
\end{aligned}\]
Consequently, we get the expression for the mixed-moment
\begin{equation}
\begin{split}
\tr{\P(\i)}&=C_{o_1-e_1-1}C_{o_2-e_2-1}C_{M-o_2-o_1+e_1+e_2}+C_{e_2-o_1}C_{o_2+o_1-e_2-e_1-2}C_{M-o_2+e_1}\\
&+\rho^2\left(C_{o_1-e_1-1}C_{M-o_1+e_1}+C_{o_2-e_1-1}C_{M-e_1+o_2-1}+C_{o_2-e_2-1}C_{M-o_2+e_2}+C_{e_2-o_1}C_{M-e_2+o_1-1}\right.\\
&\left.\qquad-2C_{o_1-e_1-1}C_{o_2-e_2-1}C_{M-o_2-o_1+e_1+e_2}-2C_{e_2-o_1}C_{o_2+o_1-e_2-e_1-2}C_{M-o_2+e_1}\right)\\
&+\rho^{4}\left(C_M-C_{o_1-e_1-1}C_{M-o_1+e_1}-C_{o_2-e_1-1}C_{M-e_1+o_2-1}-C_{o_2-e_2-1}C_{M-o_2+e_2}-C_{e_2-o_1}C_{M-e_2+o_1-1}\right.\\
&\left.\qquad +C_{o_1-e_1-1}C_{o_2-e_2-1}C_{M-o_2-o_1+e_1+e_2}+C_{e_2-o_1}C_{o_2+o_1-e_2-e_1-2}C_{M-o_2+e_1}\right),
\end{split}
\end{equation}
with 
\begin{equation}
\P(\i)=(\X\Xt)^{e_1-1}\X^2(\X\Xt)^{o_1-e_1-1}(\Xt)^2(\X\Xt)^{e_2-o_1-1}\X^2(\X\Xt)^{o_2-e_2-1}(\Xt)^2(\X\Xt)^{M-o_2}.
\end{equation}

\subsubsection{\texorpdfstring{Case $e_1<e_2<o_1<o_2$}{}}

In this case, $\Ac_{e_1,o_1}\cap\Ac_{e_2,o_o}=\emptyset$ and 
\[\Ac_{e_1,o_2}\cap\Ac_{e_2,o_1}\cong\nc_2(2(o_1-e_2-1))\times\nc_2(2(o_2-o_1+e_2-e_1-1))\times\nc_2(2(M-o_2+e_1)).\]
Consequently, we get the expression for the mixed-moment
\begin{equation}
\begin{split}
\tr{\P(\i)}&=C_{o_1-e_2-1}C_{o_2-o_1+e_2-e_1-1}C_{M-o_2+e_1}\\
&+\rho^2\left(C_{o_1-e_1-1}C_{M-o_1+e_1}+C_{o_2-e_1-1}C_{M-o_2+e_1}+C_{o_1-e_2-1}C_{M-o_1+e_2}+C_{o_2-e_2-1}C_{M-o_2+e_2}\right.\\
&\left.\qquad-2C_{o_1-e_2-1}C_{o_2-o_1+e_2-e_1-1}C_{M-o_2+e_1}\right)\\
&+\rho^{4}\left(C_M-C_{o_1-e_1-1}C_{M-o_1+e_1}-C_{o_2-e_1-1}C_{M-o_2+e_1}-C_{o_1-e_2-1}C_{M-o_1+e_2}-C_{o_2-e_2-1}C_{M-o_2+e_2}\right.\\
&\left.\qquad +C_{o_1-e_2-1}C_{o_2-o_1+e_2-e_1-1}C_{M-o_2+e_1}\right),
\end{split}
\end{equation}
with 
\begin{equation}
\P(\i)=(\X\Xt)^{e_1-1}\X^2(\X\Xt)^{e_2-e_1-1}\X^2(\X\Xt)^{o_1-e_2-1}(\Xt)^2(\X\Xt)^{o_2-o_1-1}(\Xt)^2(\X\Xt)^{M-o_2}.
\end{equation}

\subsection{\texorpdfstring{General formula in the case $k\leq M$}{}}

In the general case, the computation of $N_2(\i)$ does not yield any particular difficulty. Indeed, we have
\begin{align*}
    N_2(\i)&=\card{\Att_{e_1}(\i)\cap\Att_{e_2}(\i)}\\
    &=C_M-\sum_{2o_r-1\notin\i}(\card{\Ac_{e_1,o_r}}+\card{\Ac_{e_2,o_r}})+\sum_{\substack{2o_r-1<2o_\ell-1\\2o_r-1,2o_\ell-1\notin\i}}\left(\card{\Ac_{e_1,o_r}\cap\Ac_{e_2,o_\ell}}+\card{\Ac_{e_1,o_\ell}\cap\Ac_{e_2,o_r}}\right).
\end{align*}
The computation of $N_1(\i)$ and $N_0(\i)$ follows easily, such that 
\begin{align}
    N_O(\i)&=\sum_{\substack{2o_r-1<2o_\ell-1\\2o_r-1,2o_\ell-1\notin\i}}\left(\card{\Ac_{e_1,o_r}\cap\Ac_{e_2,o_\ell}}+\card{\Ac_{e_1,o_\ell}\cap\Ac_{e_2,o_r}}\right)\\
    N_1(\i)&=\sum_{2o_r-1\notin\i}(\card{\Ac_{e_1,o_r}}+\card{\Ac_{e_2,o_r}})-2\sum_{\substack{2o_r-1<2o_\ell-1\\2o_r-1,2o_\ell-1\notin\i}}\left(\card{\Ac_{e_1,o_r}\cap\Ac_{e_2,o_\ell}}+\card{\Ac_{e_1,o_\ell}\cap\Ac_{e_2,o_r}}\right)\\
    N_2(\i)&=C_M-\sum_{2o_r-1\notin\i}(\card{\Ac_{e_1,o_r}}+\card{\Ac_{e_2,o_r}})+\sum_{\substack{2o_r-1<2o_\ell-1\\2o_r-1,2o_\ell-1\notin\i}}\left(\card{\Ac_{e_1,o_r}\cap\Ac_{e_2,o_\ell}}+\card{\Ac_{e_1,o_\ell}\cap\Ac_{e_2,o_r}}\right).
\end{align}
Depending on the different cases listed in the previous section, we can give the explicit expression of
\[\left(\card{\Ac_{e_1,o_r}\cap\Ac_{e_2,o_\ell}}+\card{\Ac_{e_1,o_\ell}\cap\Ac_{e_2,o_r}}\right),\] for any pair $(o_r,o_\ell)$.

\newpage

\section{\texorpdfstring{Exact enumeration for $\mm{4}{8}(\rho)$}{}\label{ap:p_4_8}}

\begin{center}
    \includegraphics{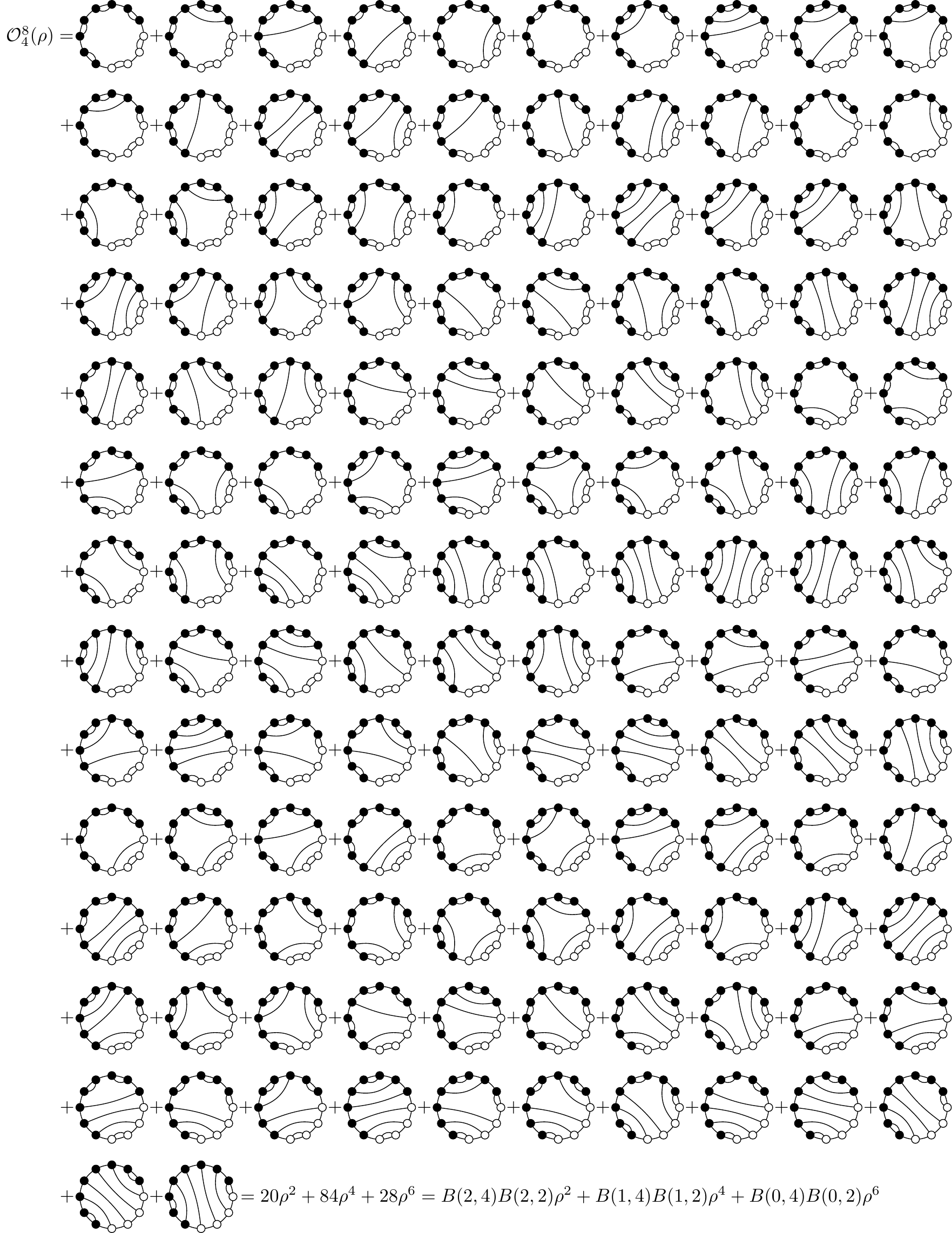}
\end{center}


\begin{thebibliography}{0}%
\makeatletter
\providecommand \@ifxundefined [1]{%
 \@ifx{#1\undefined}
}%
\providecommand \@ifnum [1]{%
 \ifnum #1\expandafter \@firstoftwo
 \else \expandafter \@secondoftwo
 \fi
}%
\providecommand \@ifx [1]{%
 \ifx #1\expandafter \@firstoftwo
 \else \expandafter \@secondoftwo
 \fi
}%
\providecommand \natexlab [1]{#1}%
\providecommand \enquote  [1]{``#1''}%
\providecommand \bibnamefont  [1]{#1}%
\providecommand \bibfnamefont [1]{#1}%
\providecommand \citenamefont [1]{#1}%
\providecommand \href@noop [0]{\@secondoftwo}%
\providecommand \href [0]{\begingroup \@sanitize@url \@href}%
\providecommand \@href[1]{\@@startlink{#1}\@@href}%
\providecommand \@@href[1]{\endgroup#1\@@endlink}%
\providecommand \@sanitize@url [0]{\catcode `\\12\catcode `\$12\catcode
  `\&12\catcode `\#12\catcode `\^12\catcode `\_12\catcode `\%12\relax}%
\providecommand \@@startlink[1]{}%
\providecommand \@@endlink[0]{}%
\providecommand \url  [0]{\begingroup\@sanitize@url \@url }%
\providecommand \@url [1]{\endgroup\@href {#1}{\urlprefix }}%
\providecommand \urlprefix  [0]{URL }%
\providecommand \Eprint [0]{\href }%
\providecommand \doibase [0]{http://dx.doi.org/}%
\providecommand \selectlanguage [0]{\@gobble}%
\providecommand \bibinfo  [0]{\@secondoftwo}%
\providecommand \bibfield  [0]{\@secondoftwo}%
\providecommand \translation [1]{[#1]}%
\providecommand \BibitemOpen [0]{}%
\providecommand \bibitemStop [0]{}%
\providecommand \bibitemNoStop [0]{.\EOS\space}%
\providecommand \EOS [0]{\spacefactor3000\relax}%
\providecommand \BibitemShut  [1]{\csname bibitem#1\endcsname}%
\let\auto@bib@innerbib\@empty
%</preamble>
\end{thebibliography}%


\begin{thebibliography}{10}

\bibitem{saxe_exact_2014}
A.~Saxe, J.~McClelland, and S.~Ganguli.
\newblock Exact solutions to the nonlinear dynamics of learning in deep linear
  neural networks.
\newblock {\em Proceedings of the International Conference on Learning
  Represenatations 2014}, 2014.
\newblock Publisher: International Conference on Learning Represenatations
  2014.

\bibitem{potters_bouchaud_2020}
Marc Potters and Jean-Philippe Bouchaud.
\newblock {\em A First Course in Random Matrix Theory: for Physicists,
  Engineers and Data Scientists}.
\newblock Cambridge University Press, 2020.

\bibitem{may1972will}
Robert~M. May.
\newblock Will a large complex system be stable?
\newblock {\em Nature}, 238(5364):413--414, aug 1972.

\bibitem{Fyodorov_Khoruzhenko_2016}
Yan~V. Fyodorov and Boris~A. Khoruzhenko.
\newblock Nonlinear analogue of the may-wigner instability transition.
\newblock {\em Proceedings of the National Academy of Sciences},
  113(25):6827--6832, Jun 2016.

\bibitem{Bizeul2020PositiveSF}
Pierre Bizeul, Maxime Clenet, and Jamal Najim.
\newblock Positive solutions for large random linear systems.
\newblock {\em ICASSP 2020 - 2020 IEEE International Conference on Acoustics,
  Speech and Signal Processing (ICASSP)}, pages 8777--8781, 2020.

\bibitem{Biroli_2018}
Giulio Biroli, Guy Bunin, and Chiara Cammarota.
\newblock {Marginally stable equilibria in critical ecosystems}.
\newblock {\em New Journal of Physics}, 20(8):83051, aug 2018.

\bibitem{anderson_introduction_2009}
Greg~W. Anderson, Alice Guionnet, and Ofer Zeitouni.
\newblock {\em An {Introduction} to {Random} {Matrices}}.
\newblock Cambridge {Studies} in {Advanced} {Mathematics}. Cambridge University
  Press, Cambridge, 2009.

\bibitem{Wigner1955CharacteristicVO}
Eugene~Paul Wigner.
\newblock Characteristic vectors of bordered matrices with infinite dimensions
  i.
\newblock {\em Annals of Mathematics}, 62:541--545, 1955.

\bibitem{wigner1967random}
Eugene~P. Wigner.
\newblock Random matrices in physics.
\newblock {\em {SIAM} Review}, 9(1):1--23, jan 1967.

\bibitem{wishart_generalised_1928}
John Wishart.
\newblock The {Generalised} {Product} {Moment} {Distribution} in {Samples} from
  a {Normal} {Multivariate} {Population}.
\newblock {\em Biometrika}, 20A(1/2):32--52, 1928.
\newblock Publisher: [Oxford University Press, Biometrika Trust].

\bibitem{ginibre_statistical_1965}
Jean Ginibre.
\newblock Statistical {Ensembles} of {Complex}, {Quaternion}, and {Real}
  {Matrices}.
\newblock {\em Journal of Mathematical Physics}, 6:440--449, March 1965.
\newblock ADS Bibcode: 1965JMP.....6..440G.

\bibitem{sommers_asymmetric}
H.~J. Sommers, A.~Crisanti, H.~Sompolinsky, and Y.~Stein.
\newblock Spectrum of large random asymmetric matrices.
\newblock {\em Phys. Rev. Lett.}, 60:1895--1898, May 1988.

\bibitem{ramli_spectral_2012}
Huda~Mohd Ramli, Eytan Katzav, and Isaac~Pérez Castillo.
\newblock Spectral properties of the {Jacobi} ensembles via the {Coulomb} gas
  approach.
\newblock {\em Journal of Physics A: Mathematical and Theoretical},
  45(46):465005, October 2012.
\newblock Publisher: IOP Publishing.

\bibitem{crisanti2012products}
A.~Crisanti, G.~Paladin, and A.~Vulpiani.
\newblock {\em Products of Random Matrices: in Statistical Physics}.
\newblock Springer Series in Solid-State Sciences. Springer Berlin Heidelberg,
  2012.

\bibitem{Dessertaine2022Cones}
Th\'eo Dessertaine and Jean-Philippe Bouchaud.
\newblock Non-self-averaging lyapunov exponent in random conewise linear
  systems.
\newblock {\em Phys. Rev. E}, 105:L052104, May 2022.

\bibitem{BrayReviewPersistence}
Alan~J. Bray, Satya~N. Majumdar, and G.~Schehr.
\newblock {Persistence and First-Passage Properties in Non-equilibrium
  Systems}.
\newblock {\em {Advances in Physics}}, 62(3):225--361, 2013.

\bibitem{nica_speicher_2006}
Alexandru Nica and Roland Speicher.
\newblock {\em Lectures on the Combinatorics of Free Probability}.
\newblock London Mathematical Society Lecture Note Series. Cambridge University
  Press, 2006.

\bibitem{bunin2016interaction}
Guy Bunin.
\newblock Interaction patterns and diversity in assembled ecological
  communities.
\newblock 2016.

\bibitem{kemp_enumeration_2011}
Todd Kemp, Karl Mahlburg, Amarpreet Rattan, and Clifford Smyth.
\newblock Enumeration of non-crossing pairings on bit strings.
\newblock {\em Journal of Combinatorial Theory, Series A}, 118(1):129--151,
  January 2011.

\bibitem{schumacher_enumeration_2013}
Paul R.~F. Schumacher and Catherine~H. Yan.
\newblock On the {Enumeration} of {Non}-{Crossing} {Pairings} of
  {Well}-{Balanced} {Binary} {Strings}.
\newblock {\em Annals of Combinatorics}, 17(2):379--391, June 2013.

\bibitem{halmagyi_mixed_2020}
Nick Halmagyi and Shailesh Lal.
\newblock Mixed {Moments} for the {Product} of {Ginibre} {Matrices}, July 2020.
\newblock arXiv:2007.10181 [cond-mat, physics:hep-th, physics:math-ph, stat].

\bibitem{Mlotkowski2010}
Wojciech Mlotkowski.
\newblock Fuss-catalan numbers in noncommutative probability.
\newblock {\em Documenta Mathematica}, 15:939--955, 2010.

\bibitem{carlitz1972sequences}
L~Carlitz.
\newblock Sequences, paths, ballot numbers.
\newblock {\em Fibonacci Quart}, 10(5):531--549, 1972.

\bibitem{adhikari2019brown}
Kartick Adhikari and Arup Bose.
\newblock Brown measure and asymptotic freeness of elliptic and related
  matrices.
\newblock {\em Random Matrices: Theory and Applications}, 8(02):1950007, 2019.

\bibitem{AIGNER199933}
Martin Aigner.
\newblock Catalan-like numbers and determinants.
\newblock {\em Journal of Combinatorial Theory, Series A}, 87(1):33--51, 1999.

\bibitem{TLAlgebras}
Georgia Benkart and Dongho Moon.
\newblock {\em Tensor product representations of Temperley-Lieb algebras and
  Chebyshev polynomials}, pages 57--80.
\newblock Rhode Island: American Mathematical Society, 2005.

\bibitem{NewellRosenblatt1962}
G.~F. Newell and M.~Rosenblatt.
\newblock {Zero Crossing Probabilities for Gaussian Stationary Processes}.
\newblock {\em The Annals of Mathematical Statistics}, 33(4):1306 -- 1313,
  1962.

\bibitem{sagemath}
{The Sage Developers}.
\newblock {\em {S}ageMath, the {S}age {M}athematics {S}oftware {S}ystem
  ({V}ersion 9.7)}, 2022.
\newblock {\tt https://www.sagemath.org}.

\end{thebibliography}
\end{document}